\documentclass{acmsiggraph}                     % final
\usepackage{booktabs} % For formal tables

\usepackage{microtype}
\usepackage{amsmath}
\usepackage{amsfonts}
\usepackage{units}
\DeclareMathAlphabet{\mathcal}{OMS}{cmsy}{m}{n}  % reset mathcal font
\usepackage{xfrac}
\usepackage[normalem]{ulem}
\usepackage{graphicx}  		% already in smrdefault
\usepackage{float}
\usepackage{url}
\usepackage{xspace}
\usepackage{color}
\usepackage{enumitem}  		% to support [resume]
\usepackage{tikz}			% support for drawing shapes
\usepackage{calc}			% support adding lengths together
\usepackage{collcell}
\usepackage{tabularx}

\usepackage{multirow}
\usepackage{eucal}
\usepackage[latin1]{inputenc}
\usepackage{listings}
\usepackage{mathrsfs}
\usepackage{subcaption}
\usepackage{cancel}
\usepackage{tcolorbox}
\usepackage{verbatimbox}

\usepackage{wrapfig}
\usepackage{ifthen}
\usepackage{algorithmicx}
\usepackage{algpseudocode}

\usepackage[ruled]{algorithm2e} % For algorithms

\SetAlFnt{\small}
\SetAlCapFnt{\small}
\SetAlCapNameFnt{\small}
\SetAlCapHSkip{0pt}

\DeclareMathOperator*{\xx}{x}

\DeclareMathOperator*{\LoeApprox}{\tilde{L}_{o \setminus e}}
\DeclareMathOperator*{\LoeApproxNew}{\tilde{L}^{new}_{o \setminus e}}

\DeclareMathOperator*{\LoApprox}{\tilde{L}_o}
\DeclareMathOperator*{\LoApproxNew}{\tilde{L}^{new}_o}

\ifnum 0 = 0

\usepackage[labelfont=bf,textfont=it]{caption}

\fi

\ifnum 1 = 0

\newcommand{\oldtext}[1]{ \textcolor{red}{\sout{#1}} }

\else

\newcommand{\oldtext}[1]{}

\fi
	
\newcommand{\picresdir}{images}

\title{Online path sampling control with progressive spatio-temporal filtering}

\author{Jacopo Pantaleoni\thanks{e-mail: jpantaleoni@nvidia.com}\\NVIDIA}


\begin{document}

\teaser{
	\includegraphics[width=180.0mm]{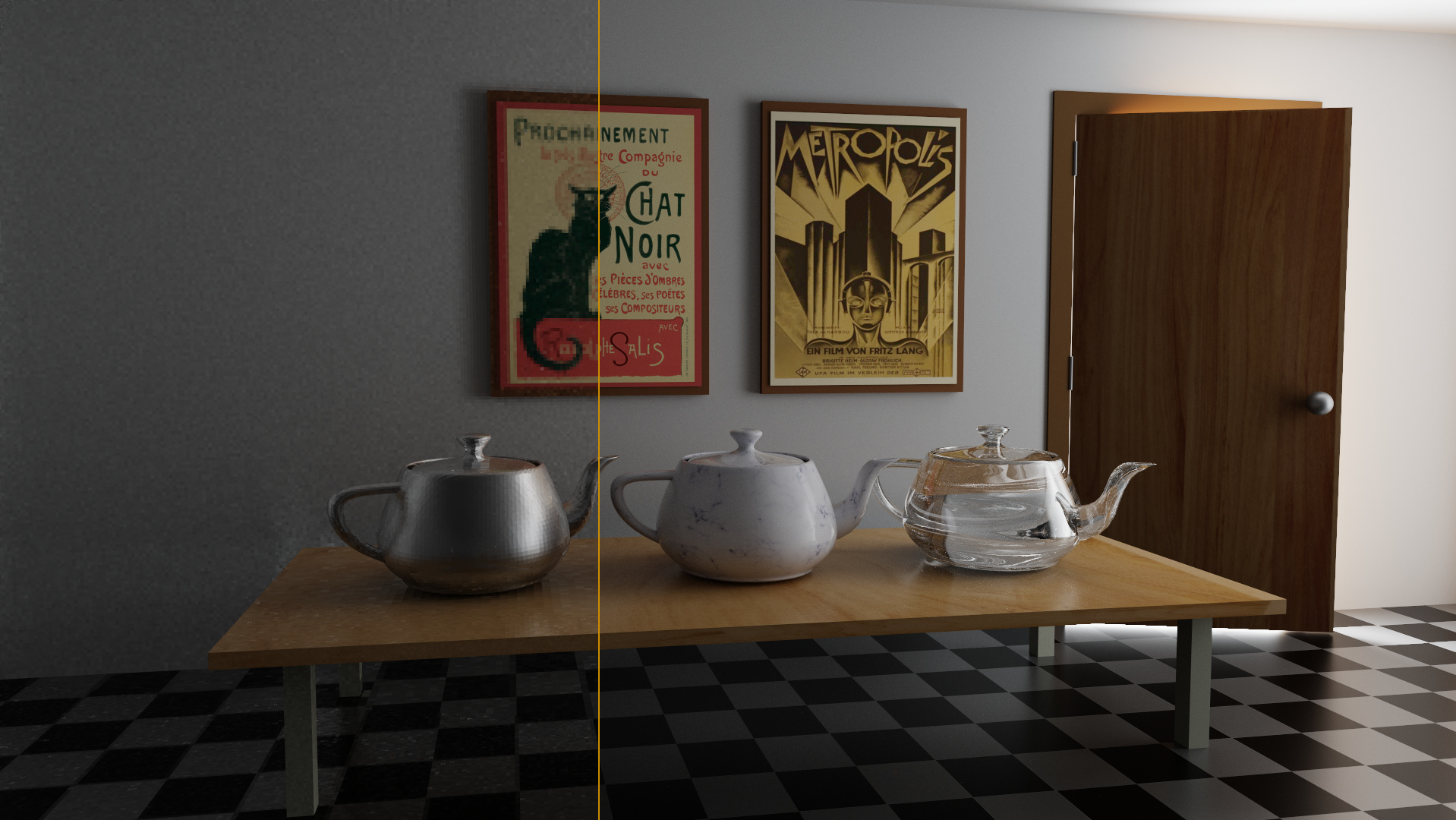}
	\caption{
		This work introduces efficient online methods to build all-frequency approximations to the light transport distribution in a scene by exploiting both its spatial and temporal coherence, and shows several ways in which these approximants can be used to control the underlying sampling process and greatly reduce sample variance in a regular path tracer.
		The left side of the picture shows the approximation built by \emph{progressive spatio-temporal filtering} after 32 iterations at 1spp, while the right side shows the converged path traced result employing this approximation as a control variate.
	}
	\label{Fig1}
}


\maketitle

\begin{abstract}
This work introduces \emph{progressive spatio-temporal filtering}, an efficient method to build all-frequency approximations to the light transport distribution into a scene by filtering individual samples produced by an underlying path sampler, using online, iterative algorithms and data-structures that exploit both the spatial and temporal coherence of the approximated light field.
Unlike previous approaches, the proposed method is both more efficient, due to its use of an iterative temporal feedback loop that massively improves convergence to a noise-free approximant, and more flexible, due to its introduction of a spatio-directional hashing representation that allows to encode directional variations like those due to glossy reflections.
We then introduce four different methods to employ the resulting approximations to \emph{control} the underlying path sampler and/or modify its associated estimator, greatly reducing its variance and enhancing its robustness to complex lighting scenarios.
The core algorithms are highly scalable and low-overhead, requiring only minor modifications to an existing path tracer.
\end{abstract}

%% ACM Computing Review (CR) categories. 
%% See <http://www.acm.org/class/1998/> for details.
%% The ``\CRcat'' command takes four arguments.

\begin{CRcatlist}
	\CRcat{I.3.2}{Graphics Systems C.2.1, C.2.4, C.3)}{Stand-alone systems};
	\CRcat{I.3.7}{Three-Dimensional Graphics and Realism}{Color,shading,shadowing, and texture}{Raytracing};
\end{CRcatlist}

%% The ``\keywordlist'' command prints out the keywords.
\keywordlist

%% The ``\copyrightspace'' command must be the first command after the 
%% start of the first section of the body of your paper. It ensures the
%% copyright space is left at the bottom of the first column on the first
%% page of your paper.

\copyrightspace

\section{Introduction}

Light transport simulation can be an arbitrarily challenging problem, due to the fact it requires to numerically estimate millions of pixel integrals whose infinite dimensional integrands may have arbitrarily high variance.
Forty years of research have produced a vast plethora of methods to increase the efficiency of this complex estimation problem, mostly based on variants of Monte Carlo integration methods, often tailored to specific scenarios.
The vast majority of these propose different strategies for \emph{path sampling}, the core operation required to numerically sample the pixel integrals. In this category fall many general purpose methods, like bidirectional path tracing and its variants \cite{Veach:PHD}, MCMC techniques like Metropolis Light Transport and its descendents \cite{Veach:1997:MLT,Kelemen:2002,Pantaleoni:2017:CML,Bitterli:2017:RJM}, as well as more ad-hoc methods such as many-lights sampling, manifold exploration \cite{Jakob:2012:ME,Kaplanyan:2014:HSLT,Hanika:2015:IHLST}, and many others.

Despite the sheer amount of research, the most popular basic method for path sampling remains path tracing, \cite{Kajiya:1986:RE}, often augmented by specific techniques to sample particular light transport events.
The reason why the most basic technique is also the most successful is to be found both in its simplicity, which leads to higher \emph{execution efficiency} on modern computing architectures, and to its very high \emph{per-sample efficiency} on \emph{average} content, that does not feature extremely complex visibility or rare events such as those due to specular-diffuse-specular transport.
In order to improve path tracing, a recent spur of research has focused the attention on \emph{path guiding}, with the idea of learning custom importance samplers on-the-fly to better guide the samples towards the more important regions of path space \cite{Vorba:2014:OLP,Herholz:2016:PIS,Muller:2017:PPG,Dahm:2017:LLT,Muller:2019:NIS}.
All of these methods can be seen as forms of online learning of different spatio-directional approximations to the underlying light field (for example based on Gaussian mixture distributions embedded in a spatial k-d tree in the approaches of Vorba et al \shortcite{Vorba:2014:OLP} and Herholz et al \shortcite{Herholz:2016:PIS}, quad-trees in the approach of Mueller et al \shortcite{Muller:2017:PPG}, simple tabulations in the approach of Dahm et al \shortcite{Dahm:2017:LLT}, and neural networks in neural importance sampling \cite{Muller:2019:NIS}).

A more limited form of online approximation of the input light field can be found in historical approaches
that cached irradiance at specific points in the scene \cite{Vray:LC,Keller:2014:PSF}.
The \emph{path space filtering} algorithm by Keller et al \shortcite{Keller:2014:PSF} constructed an approximation of the input irradiance arriving at a given vertex along a path (in the original paper, the first diffuse vertex as seen from the camera).
The approximation was built by augmenting a path tracer with a spatial data structure used to average the contributions from all paths whose first diffuse vertex happen to be close-by in space.
The averaged contributions would then be used \emph{as a replacement} for the original unbiased estimator at the specified vertex, resulting in a biased (although potentially consistent) estimator of the diffuse portion of the rendering equation.
A similar strategy is described in the documentation of v-ray's Light Cache \cite{Vray:LC}.
The path space filtering algorithm has been later extended to perform this on-the-fly using fast spatial hashing as a spatial data structure \cite{Binder:2018:FPS}.

Our work work is divided into two main parts.
In the first, we introduce a general purpose method that can be used to build similar approximations of full rank incoming and outgoing light fields, as well as their products with local brdfs, with greatly increased convergence speed.
Our methods are based on a rigorous discretization of the involved scalar light fields, and the application of efficient transport simulation methods derived from a novel combination of path-tracing and radiosity-style finite element solvers.
This part also introduces a novel spatio-directional hashing scheme allowing to compactly encode the resulting high-dimensional fields.
In the second, we study many different uses of the resulting approximants to improve the underlying path sampling estimators, not restricted to simple path guiding.
In particular, we will show that there are simpler and more efficient unbiased estimators than those used for path guiding that can be obtained by using the obtained light-field approximation as a control variate, and that by introducing some bias we can bridge the gap between unbiased estimation and biased techniques that directly use the approximation as a lighting cache.
Using control variates for path tracing had already been attempted by Lafortune and Willems \shortcite{Lafortune:1995:5D}, by employing a 5d tree based approximation of radiance built on-the-fly.
Our work shares many similarities with theirs, although we have built it on a more formal framework and faster algorithms for computing such approximations, and focused on novel algorithms and data structures geared towards a real-time implementation.

While the theoretical contributions we are introducing have general validity, our work explicitly targets real-time settings which have not been previously addressed by other path guiding methods.
Contrary to previous approaches, all of our methods are designed to be efficiently mapped to GPUs, exploit all the available parallelism and be effective even at the low sample counts typically available in real-time ray-tracing scenarios.

\ifnum 0 = 0

\section{Progressive Spatio-Temporal Filtering}

In order to describe our key algorithm, let's first consider a hypothetical discretization $\mathcal{H} = \{b_h : h \in \{1,\cdots,N\} \times \{1,\cdots,M\}\}$ of the 5-dimensional light field, seen as the tensor product of $N$ spatial basis functions and $M$ directional basis functions.
Our key insight is that we can see the construction of our desired approximation as a finite-element solver for the rendering equation using our discretization $\mathcal{H}$ as a basis.
In order to build an efficient solver, we can draw a parallel to and get inspiration from so-called \emph{radiosity} methods.
In fact, while radiosity solvers have been soon discarded in favour of the more flexible Monte Carlo methods, which proved to allow for much greater realism due to their capacity to model arbitrarily high frequency effects without the restrictions imposed by a finite-element basis, many methods developed for radiosity were nearly optimal for the finite-element setting.
In this setting, we can view the solution of the discretized rendering equation as the solution of:
\begin{eqnarray}
L_o \, &=& \, {\bf T} L_o \, + \, L_e \\
\tilde{L}_o \, &=& \, <L_o, b_h>
\end{eqnarray}
where $\bf{T}$ is the transport operator \cite{Veach:PHD}, and $<,>$ denotes projection on the basis functions.
Our approach to solving it efficiently is a hybrid between \emph{progressive radiosity} and Monte Carlo path tracing.

\subsection{Basic path tracing solver}

Since we want to obtain an \emph{online} learning algorithm that reuses the samples we generate by the underlying path sampler to build the finite element approximation, we start by considering a path-tracing based solver of the discretized equation.

The first thing to notice is that each generated path will touch as many finite-elements as it has vertices: as a consequence, we can use each sample path to update all the finite-elements it lands upon.

In the following, we will assume we may have several path sampling techniques, each associated with a sampling probability $p$ and a corresponding multiple importance sampling weight $w$ (where we omit the dependence on the technique for improved readability).
Given a sample path ${\bf x}$ with $n+1$ vertices ${\bf x} = {{\xx}}_0 {{\xx}}_2 ... {{\xx}}_{n}$, and assuming its probability and multiple importance sampling weight decompose into products of the form:
\begin{eqnarray}
p({\bf x}) &=& p({\xx}_0) \cdot p({\xx}_1|{\xx}_0) \cdots p({\xx}_{n}|{\xx}_{n-1}) \nonumber \\
w({\bf x}) &=& w({\xx}_0) \cdot w({\xx}_1|{\xx}_0) \cdots w({\xx}_{n}|{\xx}_{n-1}) \nonumber
\end{eqnarray}
we can update the solution at the finite elements touched by vertex ${\xx}_i$ using an unbiased estimator provided by the tail of the path ${\xx}_i \cdots {\xx}_{n}$.
Let's denote with $\bar{L}_o({\xx}_i,\omega_i)$ the quantity:
\begin{eqnarray}
&&\bar{L}_o({\xx}_i,\omega^o_i) \,\, = \,\, L_e({\xx}_i,\omega^o_i) \nonumber \\
&+& \sum_{j > i} L_e({\xx}_j,\omega^o_j) 
\prod_k^{i \leq k < j}
f_{k}(\omega^{i}_{k}, \omega^{o}_{k} ) G({\xx}_k, {\xx}_{k+1})
\frac{ 
	w({\xx}_{k+1}|{\xx}_k) 
} 
{ 
	p({\xx}_{k+1}|{\xx}_k) 
}, \nonumber \\
\end{eqnarray}
where $G$ denotes the geometric throughput between two vertices, $f_k$ denotes the bidirectional scattering distribution function at vertex $k$, and $\omega^i_k$ and $\omega^o_k$ denote the incoming and outgoing directions at vertex $k$.
A single-sample unbiased estimator of our approximation $\tilde{L}_{o,h} \, = \, <L_o,b_h>$ could now theoretically be obtained as:
%\begin{eqnarray}
%\tilde{L}_{o,h} \,\, &\approx& \,\, 
%\bar{L}_o({\xx}_i,\omega^o_i) b_h({\xx}_i, w^o_i) \nonumber \\
%&\cdot&
%G(x_{i-1}, x_i)
%\frac{
%	w({\xx}_0) \cdots w({\xx}_i|{\xx}_{i-1})
%}
%{
%	p({\xx}_0) \cdots p({\xx}_{i-1}|{\xx}_{i-2})
%}.
%\label{UnbiasedProjection}
%\end{eqnarray}
\begin{eqnarray}
\tilde{L}_{o,h} \,\, &\approx& \,\, 
\bar{L}_o({\xx}_i,\omega^o_i) b_h({\xx}_i, \omega^o_i) \nonumber \\
&\cdot&
\frac{
	w({\xx}_0) \cdots w({\xx}_i|{\xx}_{i-1})
}
{
	P_T({\xx}_i, \omega^o_i) 
}.
%\label{UnbiasedProjection}
\end{eqnarray}
where $P_T(x, \omega)$ is the total throughput measure probability of sampling a path which lands on the point $x$ from direction $-\omega$.
Unfortunately, as we show in the Appendix, this factor is itself a marginal probability, whose computation would involve integrating over all of path space.

However, if the spatial and angular support of the basis functions is small, and we can neglect variations inside the support, a slightly biased but practical density estimator can be obtained by shooting $N$ paths ${\bf x}_p$, with $p \in {1, \cdots N}$ and
keeping track of a weighted sum of the number of vertices $c_h$ that fall on each basis function $b_h$:
\begin{equation}
c_h = \sum_{p=1}^N \sum_i  b_h({\xx}_{p,i}, \omega^o_{p,i}) \cdot w(x_{p,i} | x_{p,i-1})
\end{equation}
and using the formula:
\begin{equation}
\tilde{L}_{o,h} \,\, \approx \,\, 
\sum_{p=1}^N 
\bar{L}_o({\xx}_{p,i},\omega^o_{p,i}) 
\frac{ 
	b_h({\xx}_{p,i}, \omega^o_{p,i}) w(x_{p,i} | x_{p,i-1})
}
{
	c_h
}.
\end{equation}

Notice how this is similar in principle to what was proposed in path space filtering \cite{Keller:2014:PSF}, except it is extended to update an approximation of the light field at all path vertices, 
using arbitrary basis functions that span both the spatial and the directional domain, and
using multiple importance sampling.
%\begin{equation}
%<L, h_{i,j}> \, = \,  \sum_{k,l} {\bf T_{i,j,k,l}} <L, h_{k,l}> + <L_e,h_{i,j}>
%\end{equation}

\subsection{Progressive solver}

In the previous section we saw how the sample paths obtained by a regular path sampler can be used to estimate the projection over the outgoing light field on a finite element basis.
The resulting method is unbiased, but has the same convergence speed as ordinary Monte Carlo path tracing.
Much faster convergence can obtained looking at solutions inspired by the radiosity literature.
Recall that the solution of the rendering equation can be written as:
\begin{equation}
L_o = L_e + {\bf T} L_e + {\bf T}^2 L_e + {\bf T}^3 L_e + \cdots
\end{equation}
In other words, the equilibrium radiance distribution is the sum of emitted radiance transported once, twice, three times, and so on.
We can exploit this fact by replacing our unbiased estimator of equation (4) with an estimator that \emph{reuses the current projection estimate} at each basis function. This is similar to the application of Jacobi iteration for the solution of a linear system, or so called progressive radiosity algorithms.
Let's call $\tilde{L}^{curr}_{o,h}$ our current estimate for the outgoing radiance projected over the basis $b_h$, and let's redefine our estimator as:
\begin{equation}
\tilde{L}^{new}_{o,h} \approx
\sum_{p=1}^N 
\bar{L}^{new}_o({\xx}_{p,i},\omega^o_{p,i})
\frac{
	b_h({\xx}_{p,i}, \omega^o_{p,i}) w(x_{p,i} | x_{p,i-1})
}
{
	c_h
}
\end{equation}
with:
\begin{eqnarray}
\bar{L}^{new}_o({\xx}_i,\omega^o_i) &=& 
L_e({\xx}_i,\omega^o_i) \nonumber \\
&+& \tilde{L}^{curr}_{o,h}({\xx}_{i+1},\omega^o_{i+1}) 
f_{i}(\omega^{i}_{i}, \omega^{o}_{i} ) G({\xx}_i, {\xx}_{i+1}) \nonumber \\
&\cdot&
\frac{ 
	w({\xx}_{i+1}|{\xx}_i) 
} 
{ 
	p({\xx}_{i+1}|{\xx}_i) 
}
\end{eqnarray}
Notice that even though this definition applies the transport operator only once, since it transports our current estimate of $\tilde{L}^{curr}_{o,h}$, its iterative application will lead to the full solution of equation (1) and (7).

In practice, in order to apply this technique, we can cast paths in waves, for example by sampling one path per pixel per frame, and performing the updates of equation (8) and (9) using the approximation corresponding to the previous frame.

%The overall structure of our update scheme is similar in spirit to that used for the Q-table creation in the reinforcement learning approach from Dahm et al \shortcite{Dahm:2017:LLT} - though here we have decoupled it from path guiding and reinforcement learning, embedding it in a general approximation framework, and extended that to represent output radiance as opposed to a pdf (the Q-table) approximating incoming radiance (particularly as the technique from Dahm et al could only handle diffuse materials), as well as explicitly keeping track of the multiple importance sampling weights.
%In the next sections, we will show further extensions to enable faster convergence, improved handling of temporal averaging, and more general handling of other fields.

\subsection{Progressive hierarchical solver}

\begin{figure}
	\includegraphics[width=84.0mm]{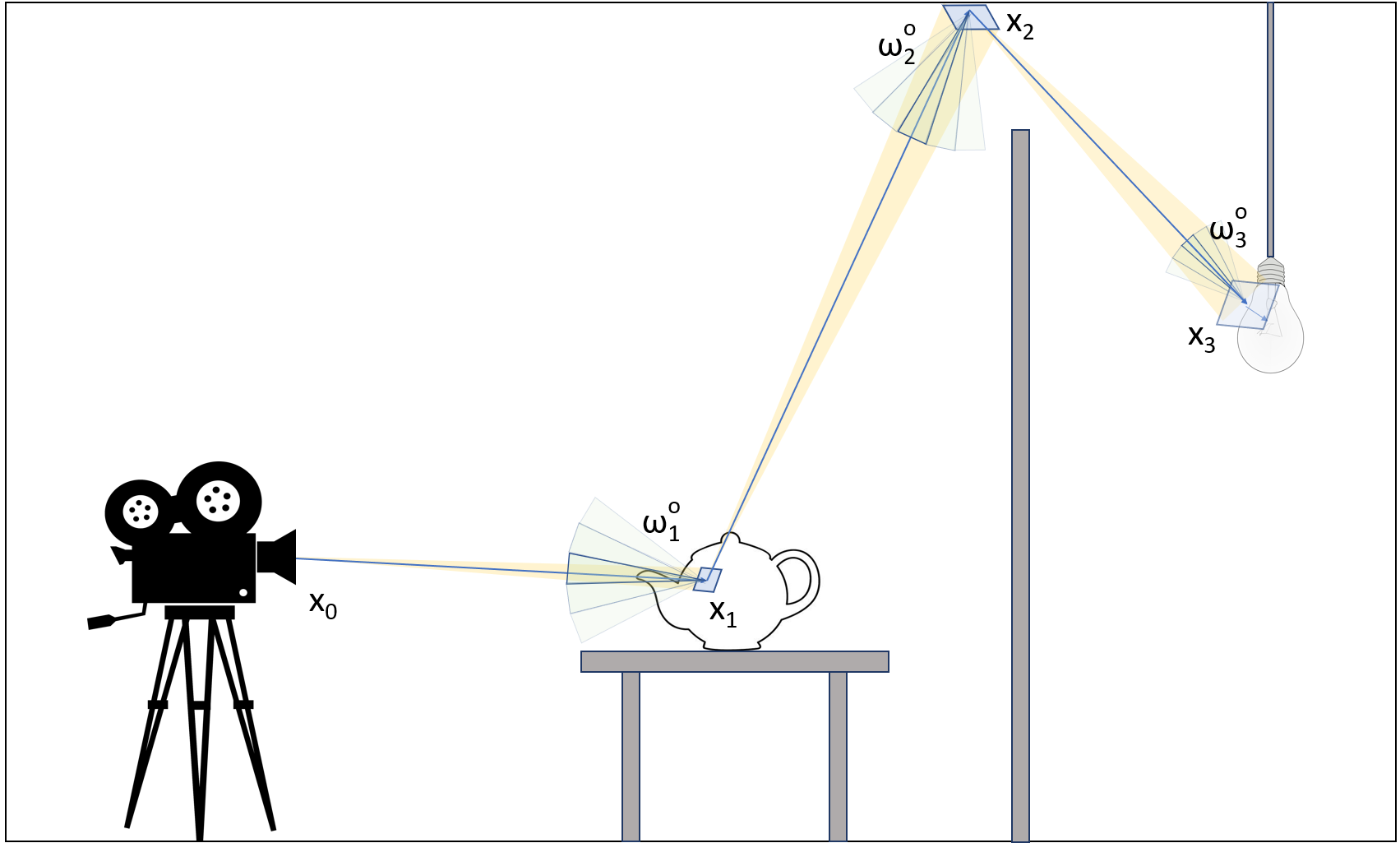}
	\caption{A schematic visualization of the path-tracing based progressive hierarchical solver: each vertex along a path touches a finite element (in this case over the outgoing radiance field), which gets updated using information at the next vertex along the same path. Approximate path footprints are used to determine the proper finite-element hierarchy level.}
	\label{PSTF-fig}
\end{figure}

The last step for obtaining even faster convergence is the use of a \emph{hierarchical} solver.
In order to do this, we have to assume a hierarchy of finite elements $\mathcal{H}_l$ where $l$ represents the level of detail.
Once we have that, we can simply replace the use of $\bar{L}^{curr}_{o,h}$ in equation (9) with a suitably selected hierarchy level $\bar{L}^{curr}_{o,h,l}$.
In our implementation, we choose the appropriate level by tracking approximate path footprints, using the heuristic described by Bekaert et al \shortcite{Bekaert:2003:CDE}.
A conceptual visualization of the final algorithm is sketched in Figure~\ref{PSTF-fig}, whereas pseudo-code for the basic path-tracing skeleton is given in Algorithm~\ref{Alg1}.
The left side of Figure~\ref{Fig1} shows an approximation built using the above algorithm.

\subsection{Temporal averaging}

In order to accomodate for dynamic scene updates, we employ a non-linear temporal averaging scheme that allows to give more weight to new samples than older ones.
In practice we do this by keeping track of two counters $c_h^{old}$ and $c_h^{new}$ for each basis function, corresponding to the cumulative counters up to the old frame, and new counters for the new frame only.
When we apply equation (8), we then use the following weighted average:
\begin{eqnarray}
&\tilde{L}^{new}_{o,h}& = \,\, (1 - \alpha^{new}) \cdot \tilde{L}^{old}_{o,h} \,\, + \nonumber \\ 
&\alpha^{new}& \cdot \sum_{p=1}^N
\bar{L}^{new}_o({\xx}_{p,i},\omega^o_{p,i})
\frac { 
	b_h({\xx}_{p,i}, w^o_{p,i}) w(x_{p,i} | x_{p,i-1})
}
{
	c_h^{new}
} \nonumber \\ 
\end{eqnarray}
where the blending coefficient $\alpha^{new}$ is computed as:
\begin{equation}
\alpha^{new} = \max \left( \sqrt{c_h^{new} / (c_h^{old} + c_h^{new})}, \, T_{max}^{-1} \right)
\end{equation}
and $T_{max}$ is a user-defined constant useful to limit the size of the temporal window.
Notice that the presence of the square root inside the blending coefficient makes the weighted average a hybrid between linear accumulation (which would be obtained without the square root), and exponential weighting, which would be obtained with a constant $\alpha^{new}$.
The counters $c_h^{old}$ can additionally be zeroed either locally, and on demand, according to custom heuristics designed to detect local changes, or globally, in the presence of large structural changes to geometry or illumination.

The reason why such a non-linear, non-exponential hybrid is desirable is to be found in the fact that, in a static setting, linear averaging corresponds to calculating the optimal Monte Carlo sample average, whereas exponential averaging gives exponentially diminishing weight to older samples, and hence discards information at an exponential rate. We found the ability to limit the loss of temporal information to be very useful, especially as some latency in the changes in indirect illumination is typically not very noticeable.

\subsection{Incoming radiance and other fields}

So far we have discussed representations that span the outgoing radiance field only, without directly encoding the incoming radiance distribution.
Some of the applications we will describe in the following sections require approximations of the product of incoming radiance and the local brdf.
Extending the representation to also account for the incoming radiance distribution would require minor modifications to the update equations. It is enough to recall that the outgoing and incoming radiance are related by:
\begin{equation}
L_i = {\bf G} L_o \nonumber
\end{equation}
where ${\bf G}$ is the propagation operator \cite{Veach:PHD}.
The basic update equations for the incoming radiance would then be:
\begin{equation}
\tilde{L}^{new}_{i,h} \approx 
\sum_{p=1}^N 
\bar{L}^{new}_i({\xx}_{p,i},\omega^i_{p,i}) 
\frac{ 
	b_h({\xx}_{p,i}, \omega^i_{p,i}) w(x_{p,i} | x_{p,i-1})
}
{
	c_h
}
\end{equation}
\begin{eqnarray}
\bar{L}^{new}_i({\xx}_i,\omega^i_i) &=&
\tilde{L}^{curr}_{o,h}({\xx}_{i+1},\omega^o_{i+1}) 
G({\xx}_i, {\xx}_{i+1}) \nonumber \\
&\cdot&
\frac{ 
	w({\xx}_{i+1}|{\xx}_i) 
} 
{ 
	p({\xx}_{i+1}|{\xx}_i) 
}
\end{eqnarray}
If we instead want to encode the product of incoming radiance and the local brdf, which we denote by $\tilde{fL}_i$, we get:
\begin{equation}
\tilde{fL}^{new}_{i,h} \approx
\sum_{p=1}^N 
\bar{fL}^{new}_i({\xx}_{p,i},\omega^i_{p,i})
\frac{ 
	b_h({\xx}_{p,i}, \omega^i_{p,i}) w(x_{p,i} | x_{p,i-1})
}
{
	c_h
}
\end{equation}
\begin{eqnarray}
\bar{fL}^{new}_i({\xx}_i,\omega^i_i) &=&
\tilde{L}^{curr}_{o,h}({\xx}_{i+1},\omega^o_{i+1}) 
\nonumber \\
&\cdot&
{ 
	f_{i}(\omega^{i}_{i}, \omega^{o}_{i} )
	G({\xx}_i, {\xx}_{i+1}) 
} 
\nonumber \\
&\cdot&
\frac{ 
	w({\xx}_{i+1}|{\xx}_i) 
} 
{ 
	p({\xx}_{i+1}|{\xx}_i) 
}
\end{eqnarray}
Yet another field that might be useful to approximate is $L_{o \setminus e} = L_o - L_e$ (corresponding to all radiance transported at least once). In order to learn the corresponding projection $\tilde{L}_{o \setminus e}$ it is enough to omit the $L_e$ term from equation (9).
Pseudo-code for tracking this field is given in Algorithm~\ref{Alg2}.
Finally, while so far we have assumed that all available sampling techniques might be used to update these fields, for path guiding applications it might in fact be beneficial to exclude some - for example because we would like to focus guided samples to areas that are not already covered by other techniques such as next-event estimation \cite{Karlik:2019:MISC}.
This would result in a down-weighted field, including only one or some of the multiple importance sampling components (and hence with weights not summing up to one).

\subsection{Basis functions and data structures: spatio-directional hashing}

The choice of basis functions and data structures is orthogonal to the methods described in this work.
However, in our implementation we have chosen a representation based on the efficient 5d spatial hashing scheme described by Binder et al \shortcite{Binder:2018:FPS}, with one crucial modification: 
for the outgoing radiance fields $\tilde{L}_o$ and $\tilde{L}_{o \setminus e}$, instead of using the surface normal at each path vertex to create a 5d hash as proposed in the original paper, we employ the outgoing direction - thus matching the representation needed to encode our 5d light field.
Similarly, for the incoming radiance field $\tilde{L}_i$ we employ a 5d hash over the position and the incoming direction.
Thus, our basis functions are essentially the product of a grid discretization of the spatial component and another grid discretization of the directional component.
This simple modification is key to keep sparsity in the encoding of our high-dimensional light fields, and what allows to work with a limited memory footprint.

In order to account for normal variation within each hash cell, we have further employed a low-order spherical harmonics representation (in our case, just two bands representing luminance in a YCoCg color decomposition) - thus making our basis functions effectively span a 7d space.

For $\tilde{fL}_i$, while we could directly adopt a 7d extension of the spatial hashing scheme encoding the position together with the incoming and outgoing directions, such a data-structure would not be practical for importance sampling due to its sparsity in the incoming directional domain.
Hence, we have also experimented combining spatial hashing for the 3d spatial component and the 2d outgoing direction with three different dense representations for the 2d incoming direction domain: regular grids, k-d trees, and spherical gaussian mixture models (GMMs).
While regular grids represent an orthonormal basis, making their update straightforward (even on parallel architectures), k-d trees and gaussian mixtures do not, and require custom update methods.

\ifnum 1 = 0

For learning GMMs, we have employed a massively parallel adaptation of the stepwise-EM algorithm described by Vorba et al \shortcite{Vorba:2014:OLP}, with the important difference that whereas Vorba et al were using each GMM to approximate the incoming radiance distribution, we use it to model the product of the incoming radiance and the brdf slice corresponding to the cone of output directions tied to a given hash cell. The details of our parallel adaptation are described in the Appendix.
For k-d trees we have experimented with entirely novel algorithms, described in the next subsection.

\fi

%which will be described in-depth in a separate work: the main idea, however, is to keep a k-d tree with a fixed number of leaves for each spatial cell, and efficiently adapt both the topology of the tree and the probabilities assigned to its leaves using statistics collected during each frame. Sampling from the k-d tree itself is performed using hierarchical sample warping \cite{Clarberg:2005:WIS}.

\subsubsection{Spherical GMMs}

Vorba et al \shortcite{Vorba:2014:OLP} proposed using spherical Gaussian mixture models to learn the incoming radiance distribution, using an algorithm dubbed stepwise-EM.
We use the same algorithm to learn the product of incoming radiance and the brdf slice tied to the given cone of output directions associated to a spatial hash cell.

The original algorithm was designed to be executed independently for each CPU thread, each working on a different GMM.
This execution model scales poorly to massively parallel GPU hardware: partly because of lack of parallelism (typically the number of cells/GMMs to update is measured in the thousands to tens of thousands per frame), partly because each GMM requires significant amounts of memory (with 6 floats per component, plus 8 more for the sufficient statistics), which cannot easily fit in on-chip memory and hence would require heavy longer-latency memory traffic.

Hence we developed two different parallel adaptations. The first and simplest is a plain SIMT adaptation that uses one SIMT lane per component.
Recall that the original algorithm is divided into two broad phases, the \emph{E}- and the \emph{M}-steps, which are executed, respectively, for every new sample and every N samples.
Focusing on a single GMM, given the sufficient statistics $u_i$ at step $i$, represented as a matrix with $C$ rows and 8 columns, where $C$ is the number of mixture components, and a new sample $s_i = (s_x, s_y)$ with weight $w_i$, the \emph{E}-step can be written as:
\begin{equation}
u_i[c] = a_i \cdot u_{i-1}[c] + b_i \cdot v_i
\label{SufficientStatsEq}
\end{equation}
where:
\begin{eqnarray}
v_i &=& (1, s_x, s_y, s_x \cdot s_x, s_y \cdot s_y, s_x \cdot s_y, 1 / \gamma_c, 1 / b_i) \nonumber \\
a_i &=& (1 - i^{-\alpha}) \nonumber \\
b_i &=& i^{-\alpha} \cdot w_i \cdot \gamma_c
\end{eqnarray}
and $\gamma_c$ is the responsibility of the $c$-th component of the current GMM for the point $s_i$:
\begin{equation}
\gamma_c = \frac{p_c(s_i)}{\sum_j p_j(s_i)}
\end{equation}

Notice how the update equations (\ref{SufficientStatsEq}) can be trivially parallelized across components; the only computation that needs special care is the calculation of $\gamma_c$, which requires all threads to participate in the computation of the denominator, essentially computing a parallel reduction.

This one-thread-per-component mapping is significantly faster than the trivial one-thread-per-GMM mapping, especially as it allows each thread to only keep one component worth of data in registers.
However, load-balancing might still be an issue, as some hash cells/GMMs might receive many more samples than others, requiring an uneven number of $E$-step iterations.

Hence, we devised an even broader parallelization strategy that uses one thread per sample.
In fact, while the recurrent relation (\ref{SufficientStatsEq}) seems to impose complex dependencies that do not allow parallelizing across samples, expanding the recurrent relation one can obtain:
\begin{equation}
u_i[c] = b_i \cdot v_i + \sum_{1 \leq j < i} b_{j} \cdot v_{j} \cdot\prod_{j < k \leq i} (1 - k^{-\alpha})
\label{UnrolledSufficientStatsEq}
\end{equation}
that is to say: the sufficient statistics for the i-th sample are obtained by summing up the contributions $v_j$ of all the samples preceding it, weighted by a product term of the form:
\begin{equation}
g(j) = \prod_{j < k \leq i} (1 - k^{-\alpha})
\end{equation}
The main observation behind our parallelization strategy is that we can compute the logarithm of $g(j)$ with a parallel prefix-sum:
\begin{equation}
log(g(j)) = \sum_{j < k \leq i} (1 - k^{-\alpha})
\end{equation}
Hence, we first sort the samples recorded during a frame by GMM; then we proceed by calculating the coefficients $log(g(j))$ for each sample contributing to each GMM with a segmented parallel prefix-sum (where each GMM defines a segment) and finally, we evaluate equation (\ref{UnrolledSufficientStatsEq}) using another segmented parallel reduction.

\subsubsection{Spherical k-d trees}

In order to develop scalable algorithms to efficiently access and update tens of thousands of spherical k-d trees in a massively parallel setting, we opted for a simple and compact representation, constraining each k-d tree to possess the same number of leaves, while freely adapting both their topology and the sampling probabilities assigned to their leaves using statistics collected during each frame.

Constraining each tree to possess the same number of leaves $L$, and consequently the same number of nodes $2 \cdot L - 1$, allows to store the trees compactly in deterministic order and avoids random memory allocation, striking a careful balance between memory-access efficiency and representational flexibility.

Our k-d trees span a 2d domain $[0,1]^2$, representing a uniform parameterization of the sphere.
Starting from uniformly split k-d trees (essentially representing uniform grids), we update the probability of each leaf according to the sum of the contributions of the samples falling within it - so as to keep the sampling probability of each leaf proportional to the integral of the incoming radiance times the BRDF (remembering that each k-d tree is tied to a given cone of output directions).

Finally, we update each tree's topology using the \emph{split-collapse} algorithm employed by Pantaleoni for reinforcement light-cuts learning \shortcite{Pantaleoni:2019:RLL}: at the end of each frame, for each k-d tree we look at the leaf with highest probability $l_{max}$, and the parent with lowest probability $p_{min}$: if the probability of the former is higher than that of the latter times a constant $T$, i.e. $P(l_{max}) > T \cdot P(p_{min})$, we split $l_{max}$ and collapse $p_{min}$.
Similarly to the original implementation of \emph{split-collapse}, we launch one thread block per k-d tree, and parallelize all phases of the algorithm.

During rendering, sampling from each k-d tree itself is performed using hierarchical sample warping \cite{Clarberg:2005:WIS}. For each sample, we store its primary sample space coordinate together with its MIS weighted contribution in order to update the leaf probabilities at the end of the frame.

\section{Relation to previous work}

Besides the similarities and the differences to path space filtering \cite{Keller:2014:PSF,Binder:2018:FPS} already mentioned at the end of section 2.1, the overall structure of our update scheme shares some similarities to that used for the Q-table creation in the reinforcement learning approach from Dahm et al \shortcite{Dahm:2017:LLT}. Here, however, the creation of all our approximators is entirely decoupled from path guiding and reinforcement learning and simply embedded in a more general approximation framework tied to arbitrary path sampling schemes, and extended to represent outgoing radiance, incoming radiance and the product of incoming radiance with the local brdf, as opposed to a pdf (the Q-table) approximating incoming radiance only (limiting the technique from Dahm et al \shortcite{Dahm:2017:LLT} to only handle path guiding for diffuse materials); this new, more flexible framework is further extended to explicitly keep track of multiple importance sampling and weighted distributions.

Furthermore, we have shown how to enable faster convergence by using hierarchical basis functions (Section 2.3) and an improved, non-exponential handling of temporal averaging (Section 2.4), and we have extended the scope of practical implementations to use a larger set of basis functions that can span the complete 7d field needed to represent product distributions while using acceptable storage. Key to the latter is the use of a representation that is sparse in the outgoing direction, achieved by modifying the sparse spatial hashing scheme of Binder et al \shortcite{Binder:2018:FPS} to hash over the outgoing direction as opposed to using the surface normal. Detail due to normal variation in each cell is instead optionally recaptured using the YCoCg spherical harmonics representation.

To our knowledge our framework is the first that can directly handle product distributions without computing the product of separate approximations of the incoming light field and the local brdf on-the-fly, as done by Herholz et al \shortcite{Herholz:2016:PIS}, an operation that is rather expensive and that requires brdf representations that can easily be converted to the target basis functions (again limiting the applicability to complex material models).

Finally, we have provided novel scalable algorithms for efficiently learning GMMs and adaptive k-d trees on massively parallel architectures.

In the next sections we will further show how the resulting approximations can be used to enable a new set of estimators and control methods that go beyond simple path guiding.

\section{Unbiased estimation}

Once we have an approximation of the outgoing and the incoming light fields $\tilde{L}_{o}$ and $\tilde{L}_{i}$ we can directly use them to \emph{control} our path sampling estimators.
While previous research on path guiding methods has already covered using similar approximations for importance sampling, we will show how they can also be very effectively employed as \emph{control variates}.

\subsection{Importance sampling (or path guiding)}

All path guiding methods are based on importance sampling from either an approximate representation of the incoming radiance distribution \cite{Vorba:2014:OLP,Muller:2017:PPG,Dahm:2017:LLT} or a representation of the product of incoming radiance and the local brdf \cite{Herholz:2016:PIS} that is learnt on-the-fly. Progressive spatio-temporal filtering allows to build exactly such a representation $\tilde{fL}_{i}$.
In order to make the process unbiased, during each frame we importance sample $\tilde{fL}^{old}_{i}$ while updating an entirely separate approximation $\tilde{fL}^{new}_{i}$ that is only going to be used in the next frame.

In the following it will be convenient to look at local path sampling as a recursive solution of the rendering equation written in its integral form:
\begin{equation}
L_o({\xx}, \omega^o) = L_e({\xx}, \omega^o) + \int_{\Omega} L_i({\xx}, \omega^i) f_x(\omega^i,\omega^o) cos(\theta^i)d\omega^i
\label{RecursiveRenderingEq}
\end{equation}

Given a path vertex $x_j$ and an output direction $\omega^o_j$, we solve for $L_o({\xx}_j, \omega^o_j)$ by sampling a direction $\omega^i_j$ according to some projected solid angle probability $p^\perp(\omega^i_j|x_j)$ and using the single-sample estimator:
\begin{equation}
L_{o \setminus e}({\xx}_j, \omega^o_j) \approx 
L_i({\xx}_j, \omega^i) f_{x_j}(\omega^i_j,\omega^o_j) 
\frac{ 
	w(\omega^i_j|x_j)
}
{
	p^\perp(\omega^i_j|x_j)
}
\end{equation}
This view makes it clear that the changes due to the approximation-based importance sampling technique are simply embedded in the vertex sampling probabilities $p({\xx}_{j+1}|{\xx}_j) = p^\perp(\omega^i_j|x_j)  G({\xx}_j, {\xx}_{j+1})$, and do not change the form of the final path sampling estimators.

%Given a complete sample path ${\bf x} = {\xx}_0 {\xx}_2 ... {\xx}_{n}$ joining the camera with a light source, the corresponding path tracing estimator for the $j$-th pixel is the usual one:
%\begin{equation}
%I_j \approx
%\frac{	
%	m_j({\bf x}) w({\bf x})	
%}
%{
%	p({\bf x})	
%} 
%\end{equation}
%where $m_j$ is the measurement contribution function:
%\begin{eqnarray}
%m_j({\bf x}) &=&
%L_e(x_n, \omega^o_n) \nonumber \\
%&\cdot& \prod_{i=1}^{n-1} \big[ 
%f_i(\omega^i_i,\omega^o_i) G(x_i,  x_{i+1})
%\big] \nonumber \\
%&\cdot& W_e^j(x_0, x_1)
%\end{eqnarray}
%In this case, the changes due to the approximation-based importance sampling technique are simply embedded in the vertex sampling probabilities $p({\bf x}) = p({\xx}_0) \cdot p({\xx}_1|{\xx}_0) \cdots \cdot p({\xx}_{n}|{\xx}_{n-1})$.

In practice, at each path vertex we combine sampling according to $\tilde{fL}_{i}$ with a defensive strategy based on the BSDF by means of multiple importance sampling.
Similarly, other vertex sampling techniques such as next-event estimation can be easily incorporated.

\subsection{Control variates}

As anticipated, another estimator can be obtained using our new approximations as a control variate.
Suppose we are integrating a function $g({\bf x})$, and have another function $h({\bf x})$ with known integral $I_h$; we can then obtain an unbiased estimator of the integral of $g$ as:
\begin{equation}
E[g] \approx
\frac{	
	\left[ g({\bf x}) - \beta h({\bf x}) + \beta I_h \right]
}
{
	p({\bf x})	
}.
\end{equation}
where $\beta$ is a control parameter. The function $h$ is said to be a control variate \cite{Owen:2000:SAE}.

Again, by recalling that at each part vertex $({\xx}_j, \omega^o_j)$ we are locally solving equation (\ref{RecursiveRenderingEq}),
we can exploit this fact by using the control variate $h = \tilde{f L_i}({\xx}_j, \omega^i_j)$ with known integral $\tilde{L}_{o \setminus e}({\xx}_j, \omega^o_j)$.
The corresponding estimator will be:
\begin{eqnarray}
\label{CVEstimator}
&&L_{o \setminus e}({\xx}_j, \omega^o_j) \approx  \nonumber \\
&& \left[ L_i({\xx}_j, \omega^i) f_{x_j}(\omega^i_j,\omega^o_j) - \beta \tilde{f L_i}({{\xx}_j}, \omega^i_j) + \beta \tilde{L}_{o \setminus e}({\xx}_j, \omega^o_j) \right] \cdot \nonumber \\
&& 
\cdot
\frac{ 
	w(\omega^i_j|x_j)
}
{
	p^\perp(\omega^i_j|x_j)
}
\end{eqnarray}

As we will show in the results section, such a control variate can be surprisingly effective.
In practice, we have also observed that it is sufficient and sometimes beneficial to restrict its application to the first few vertices along a path.
Moreover, while for a given function $h$ optimal variance reduction would require optimizing for $\beta$, we have obtained excellent results even with fixed $\beta$ in the range $[0.5,1]$.

\subsection{Importance sampling with control variates}

As shown by He and Owen \shortcite{He:2014:OMW} combining importance sampling and control variates can theorically lead to even lower variance estimators.
For several importance sampling techniques $p_1, \cdots, p_m$, He and Owen suggest using the following estimator:
\begin{equation}
E[g] \approx
\frac{	
	\left[ g({\bf x}) - \beta^T h({\bf x}) + I_h \right]
}
{
	p_\alpha({\bf x})	
} 
\label{MISCV}
\end{equation}
where $p_\alpha$ is a weighted average of the probabilities $p_\alpha = \sum_i \alpha_i p_i$, $h$ is a vector function $h = (p_1({\bf x}), \cdots, p_m({\bf x}))$ and $\alpha$ and $\beta$ are multi-dimensional parameters in $\mathbb{R}^m$.

In our case, since at each vertex we use both the BSDF defensive strategy, $p_0$, and the approximation-based importance sampling strategy $p_1 = \tilde{f L_i}$, we have chosen to employ a simpler combined control variate of the form:
\begin{equation}
E[g] \approx
\frac{	
	g({\bf x}) - \beta \left[ p_1({\bf x}) - p_0({\bf x}) \right]
}
{
	p_\alpha({\bf x})
} 
\end{equation}
where $\alpha = (\alpha_0,\alpha_1)$ is the ratio of samples allocated to each of the two strategies, and $\beta$ is again a simple scalar control.
The choice of the scalar function $p_1 - p_0$ as a control variate follows the approach described by Li et al \shortcite{Li:2013:TIS}, and it is equally efficient as the original estimator (\ref{MISCV}), which is singular in $\beta$. Moreover, it has the advantage of having a zero integral.

\section{Biased estimation}

In the previous section we have covered many alternative unbiased estimators of the rendering equation that can be built on top of our online approximation of the underlying light field.
In this section we'll cover yet another biased estimator that can further reduce the overall error at the cost of some bias.
The basic idea is to take the control variate estimator (\ref{CVEstimator}) with $\beta = 1$, and reparameterize it as:
\begin{eqnarray}
\label{BiasedEstimator}
&&L_{o \setminus e}({\xx}_j, \omega^o_j) \approx  \nonumber \\
&& \cdot 
\left[ 
\gamma 
\left(
L_i({\xx}_j, \omega^i) f_{x_j}(\omega^i_j,\omega^o_j)
- \tilde{f L_i}({{\xx}_j}, \omega^i_j)
\right)
+ \tilde{L}_{o \setminus e}({\xx}_j, \omega^o_j) \right] \cdot \nonumber \\
&& 
\cdot
\frac{ 
	w(\omega^i_j|x_j)
}
{
	p^\perp(\omega^i_j|x_j)
}
\end{eqnarray}
We look at this estimator as a predictor-corrector model, where the $\tilde{L}_{o \setminus e}({\xx}_j, \omega^o_j)$ term plays the role of the predictor, and the difference:
\begin{equation}
\gamma \left( L_i({\xx}_j, \omega^i) f_{x_j}(\omega^i_j,\omega^o_j) - \tilde{f L_i}({{\xx}_j}, \omega^i_j) \right) \nonumber
\end{equation}
plays the role of the corrector. By setting $\gamma < 1$ we simply bias the solution towards the predictor.
Notice that, again, we can use such an estimator at any (or even every) vertex along a path.
By setting $\gamma = 0$ and using it only at the very first diffuse vertex, we can reproduce the effect of the path space filtering algorithm by Keller et al \shortcite{Keller:2014:PSF}, except it is extended to handle all-frequency lighting and use the more efficient progressive spatio-temporal filtering approximation.

\section{Results and discussion}

In order to test the various algorithms we have used a reproduction of Eric Veach's \emph{door ajar} scene, notoriously designed to be hard for light transport simulation. We have also slightly modified the scene to make use of more modern rendering features, such as layered material models that are not either simply diffuse or purely specular.
Figure~\ref{Fig1} shows a perfectly converged path traced rendering using the control variate estimator (on the right), together with the approximation of the outgoing light field $\tilde{L}_o$ that we obtain in a few iterations (32 frames at 1spp each) with our progressive spatio-temporal filtering algorithm.
Notice how our progressive solver converges extremely quickly, and can reproduce high frequency transport such as that due to glossy interreflections.

We implemented all the above methods on top of the highly efficient massively parallel path tracing kernels in Fermat, paying attention to exploit all available parallelism and reducing synchronization overheads in all phases of computation, including updates of the GMM and k-d tree representations.
The control-variate and biased approaches result in very low-overhead: roughly 2-4ms on top of a total per-frame cost of 16-20ms for path-tracing a 1080p image at 1spp on a system equipped with a single RTX 2080 Ti.
The path guiding approaches resulted in much higher overheads despite all our efforts: importance sampling and evaluating the importance sampling pdf of the GMM, together with their updates, resulted in an added cost of about 70ms per frame. Our custom k-d tree method resulted in a lower overhead of about 30-40ms per frame.
The path guiding kernels also required a much higher memory consumption: while the pure spatial hashing representations used for the control variates requires only 48 bytes per cell (24 for the biased estimator with $\gamma = 0$), and a hash table of 4M entries,
the GMM representation requires 6 floats per component per spatial cell, resulting in 96 bytes per cell for a 4-component GMM and 182 bytes per cell for an 8-component one.
Our adaptively split k-d trees require exactly 128 bytes per spatial cell.

Table~\ref{PSTF-comparisons} shows a comparison of all our estimators from sections (3) and (4) for various sample counts, as well as pure path tracing with and without next-event estimation.
In particular, we compare the PSTF-based pure path guiding / importance sampling estimator (PSTF-IS), the control variate estimator (PSTF-CV), the combined importance sampling + control variate estimator (PSTF-IS-CV) and finally the biased estimator with $\gamma = 0$ applied at the \emph{second} vertex along a path (PSTF-B).
All the PSTF-based estimators include next-event estimation.
For the 1spp images, we used a warm-up phase of 32 frames to obtain nearly converged approximations.
Notice how at 1spp the control variate based estimators provide for widely improved convergence even compared to the path guiding estimator (despite being far cheaper to evaluate), and gets orders of magnitude lower variance than the raw path sampling estimator.
The biased estimator improves the results even further. Notice that applying it at the second bounce helps to reduce its bias and to hide some of its associated lower frequency noise.

\begin{figure}
	\includegraphics[width=85.0mm]{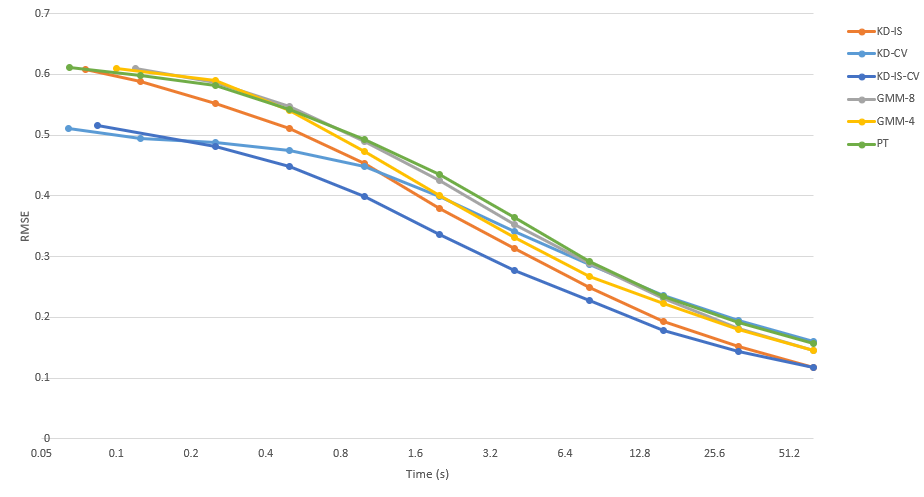}
	\caption{A graph of the root mean square error (RMSE) plotted over time for all our unbiased estimators.}
	\label{VarianceAnalysis}
\end{figure}

Figure~\ref{VarianceAnalysis} compares the performance by graphing over time the root mean square error (RMSE) of all our unbiased estimators. The comparison, obtained measuring averages over a variety of scenes, includes the k-d tree based importance-sampling estimator (KD-IS), the k-d tree based control variate estimator (KD-CV),
the k-d tree based coupled importance-sampling plus control variate estimator (KD-IS-CV), the GMM based importance-sampling estimator with 4 components (GMM-4), the GMM based importance-sampling estimator with 8 components (GMM-8), and plain path-tracing with next-event estimation (PT). 
As our work is strictly addressing real-time rendering scenarios, we explicitly opted not to compare against previous path-guiding methods that were designed solely for offline rendering and had much larger overheads and parallelization bottlenecks.
Notice how the control variate k-d tree based estimator at real-time settings (60ms) is roughly 13x faster than a pure path-tracing estimator (i.e. the path-tracing estimator has the same variance after ~0.8s), and between 10x and 13x times faster than the GMM based path-guiding estimators.
Path guiding based approaches start becoming more efficient than the pure control variate estimator after 0.8-2 seconds depending on the implementation, whereas the combined importance-sampling plus control variate estimator starts providing an advantage after 0.2 seconds, although the advantage brought in by the control variate is asymptotically lost.

Figure~\ref{Staircase2} shows an equal-time comparison on a different scene where the biased estimator provides much less noisy results than simple path tracing, despite the much simpler light transport configuration.
Figure~\ref{ArchInterior} shows another equal-time comparison on a complex visibility scene where our biased estimator is compared to plain path-space filtering.

We also want to highlight that while other works on path guiding have overlooked this aspect, in order to make such comparisons meaningful it is \emph{absolutely critical} to include next-event estimation, since this seemingly basic technique can reduce variance thousand-folds in the presence of complex lighting, and is by itself way more effective than path guiding alone\footnote{Except for caustics, where simple next-event estimation cannot work, and one should resort to more complex approaches \cite{Hanika:2015:MNE}}.

Finally, we speculate that pure path guiding / importance sampling may not be very cost-effective when compared to a path-tracer with next-event estimation due to the fact that imperfect importance sampling techniques can lead to higher variance samples, that can only be partially mitigated by multiple importance sampling.
While Owen and Zhou \cite{Owen:2000:SAE} have discussed safer strategies to minimize the additional variance due to locally suboptimal importance sampling techniques, these involve costly convex optimization of the multiple importance sampling weights.
Moreover, it is important to notice that in many cases path guiding only helps with a tiny portion of a path: for example, in the \emph{door ajar} scene it helps going through the door. However, both inside the first room and once in the back room, where multiple bounces among the walls diffuse out the overall lighting, path guiding cannot provide much help - though it still adds its associated overheads.
Interestingly, Vorba et al \shortcite{vorba19guiding} have recently introduced a simpler online MIS optimization technique performing stochastic gradient descent on KL divergence that seems to minimize the negative impact of suboptimal importance sampling decisions by reducing their probability\footnote{as opposed to introducing a control variate as in \cite{Owen:2000:SAE}.}. While we did not have a chance to test this yet, even this simple algorithm is bound to increase memory traffic and incur additional synchronization overhead, as it requires the use of spin-locks that are potentially expensive in a massively parallel scenario. 

\begin{figure}
	\includegraphics[width=85.0mm]{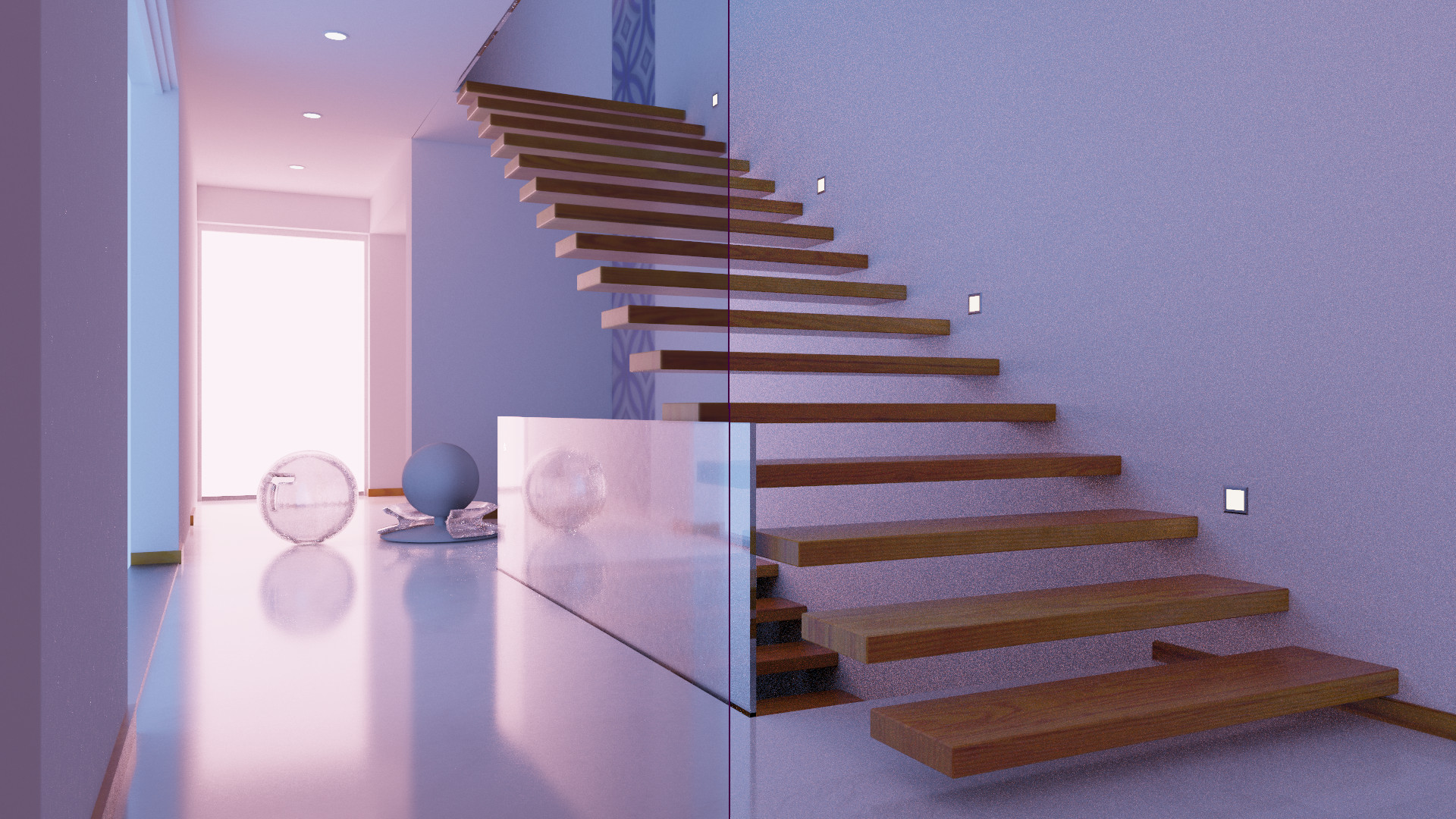}
	\caption{An equal-time comparison of our biased estimator performed at the first vertex (left side) against path tracing with next-event estimation (right side).}
	\label{Staircase2}
\end{figure}

\subsection{Future work}

Besides applying the same techniques to the handling of participating media, a natural extension of this work would be to combine it with adjoint-driven Russian roulette and splitting \cite{Vorba:2016:ARR}, which might help further reduce variance at a low additional cost.
Another venue would be to couple it with progressive photon mapping \cite{Hachisuka:2008:PPM,Hachisuka:2008:SPPM} or vertex merging techniques \cite{Hachisuka:2012}.
Yet another potential area is coupling it with the latest results on improved multiple importance sampling 
\cite{Kondapaneni:2019:OMIS,Karlik:2019:MISC,Sbert:2019:GMIS}.
Finally, it would be interesting to automate the choice of the constant $\gamma$ in our biased estimator to minimize total error, seen as the sum of bias and variance.

\begin{table*}
	
	\addtolength{\tabcolsep}{-4pt}
	\begin{tabularx}{\textwidth}{|c|c|c|c|}
		\hline
		&
		1 spp &
		32 spp &
		60s
		\\ \hline
		\rotatebox{90}{PT w/o NEE} &
		% 0.626026
		\multicolumn{1}{c}{\includegraphics[width=56.0mm]{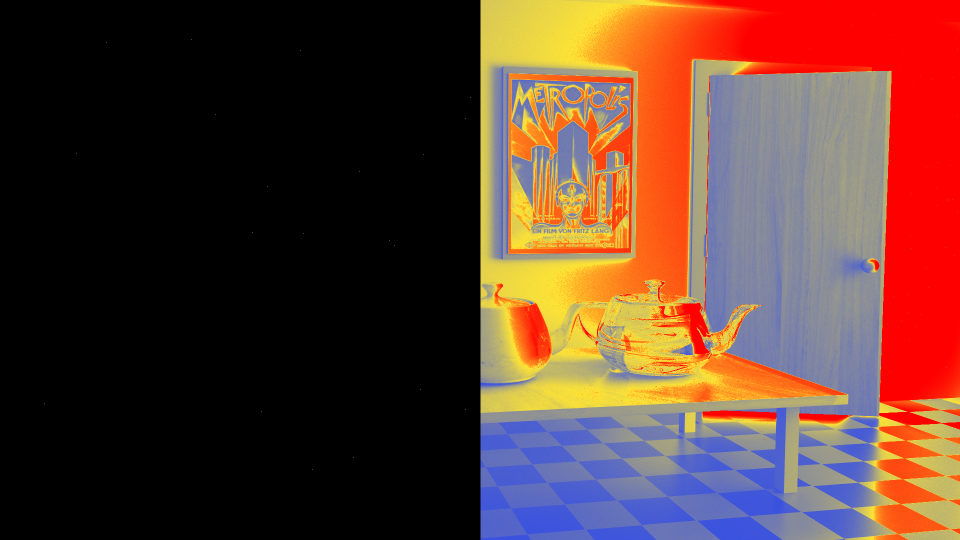}} &
		% 0.625053
		\multicolumn{1}{c}{\includegraphics[width=56.0mm]{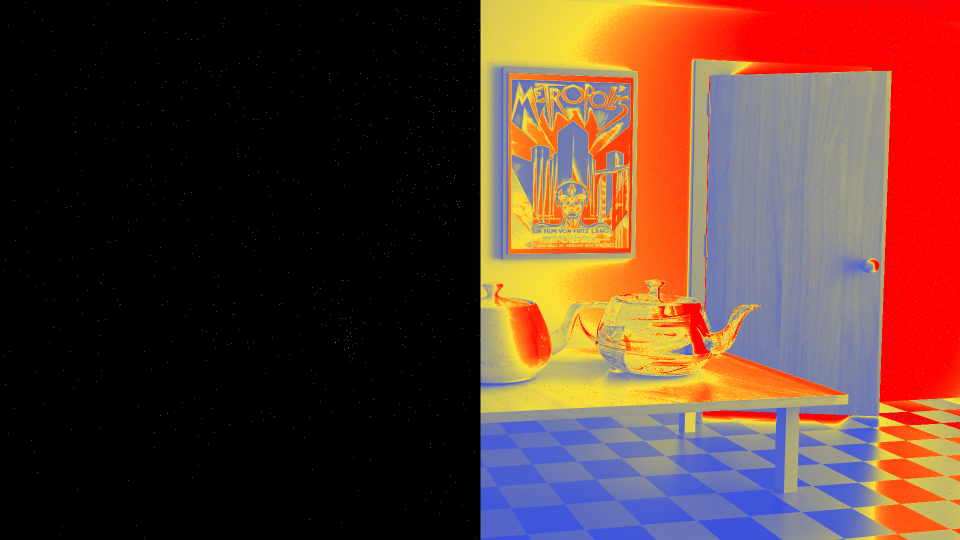}} &
		% 0.564618
		\multicolumn{1}{c}{\includegraphics[width=56.0mm]{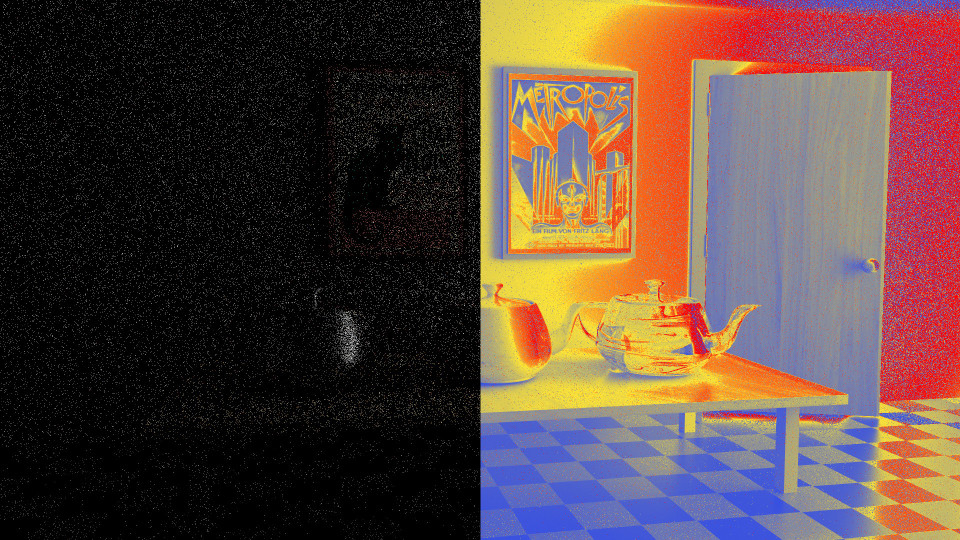}}
		\\ \hline
		&
		\tiny{RMSE 0.625053} &
		\tiny{RMSE 0.625053} &
		\tiny{RMSE 0.564618}
		\\ \hline
		\rotatebox{90}{PT w/ NEE} &
		% 0.618978
		\multicolumn{1}{c}{\includegraphics[width=56.0mm]{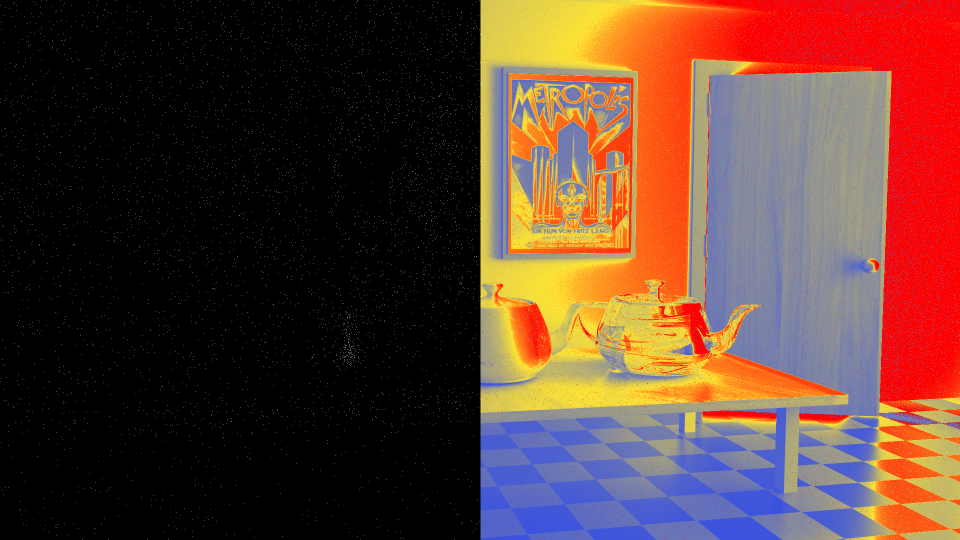}} &
		% 0.495649
		\multicolumn{1}{c}{\includegraphics[width=56.0mm]{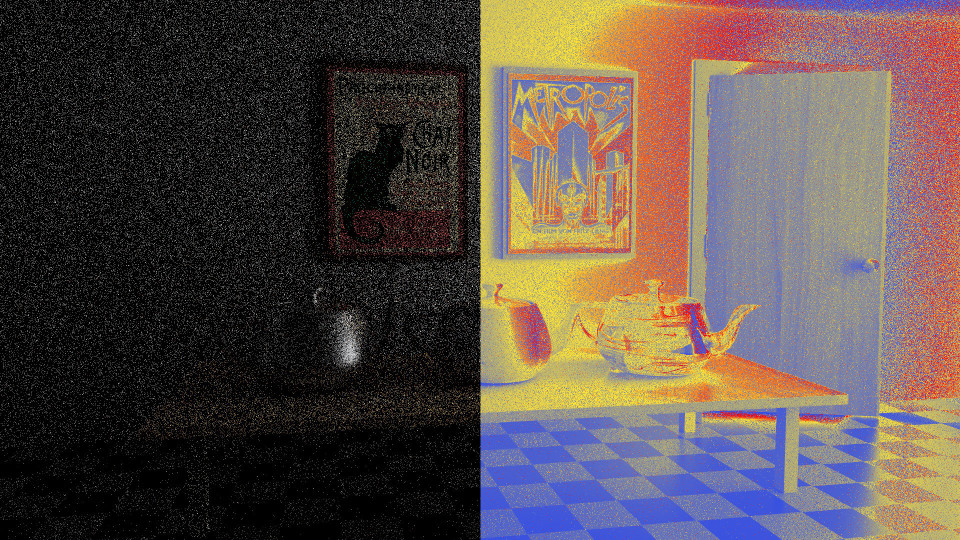}} & 			
		% 0.155520
		\multicolumn{1}{c}{\includegraphics[width=56.0mm]{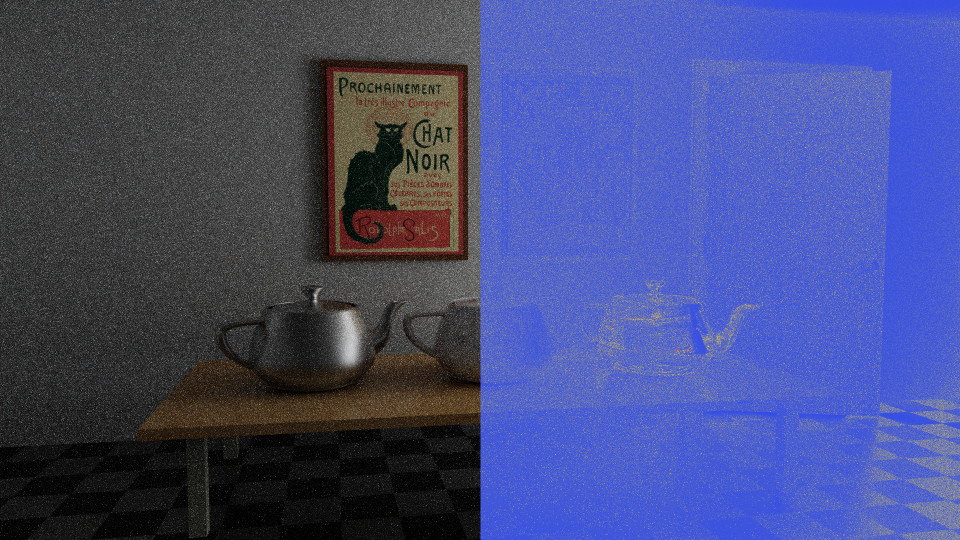}}			
		\\ \hline
		&
		\tiny{RMSE 0.618978} &
		\tiny{RMSE 0.495649} &
		\tiny{RMSE 0.155520}
		\\ \hline
		\rotatebox{90}{PSTF-IS} &
		% 0.608228	
		\multicolumn{1}{c}{\includegraphics[width=56.0mm]{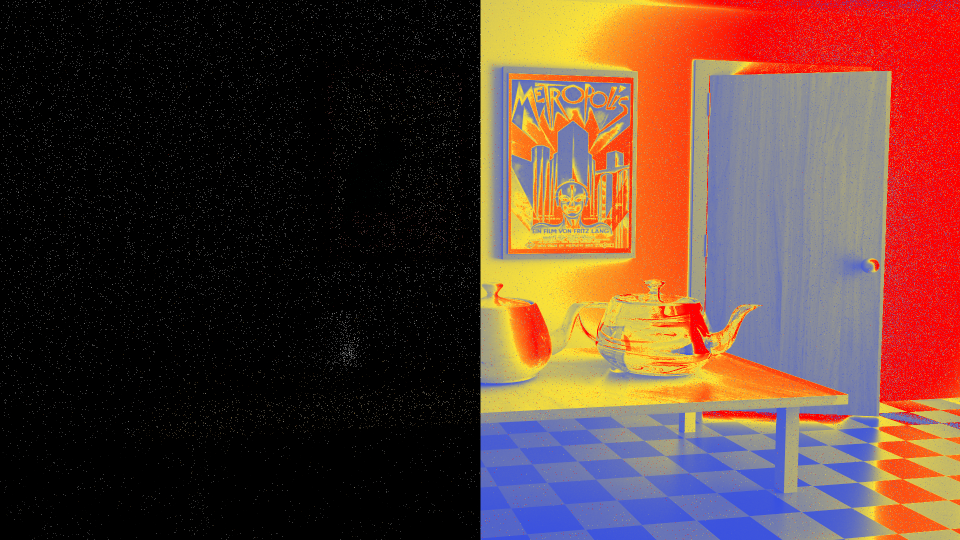}} & 	
		% 0.424356	
		\multicolumn{1}{c}{\includegraphics[width=56.0mm]{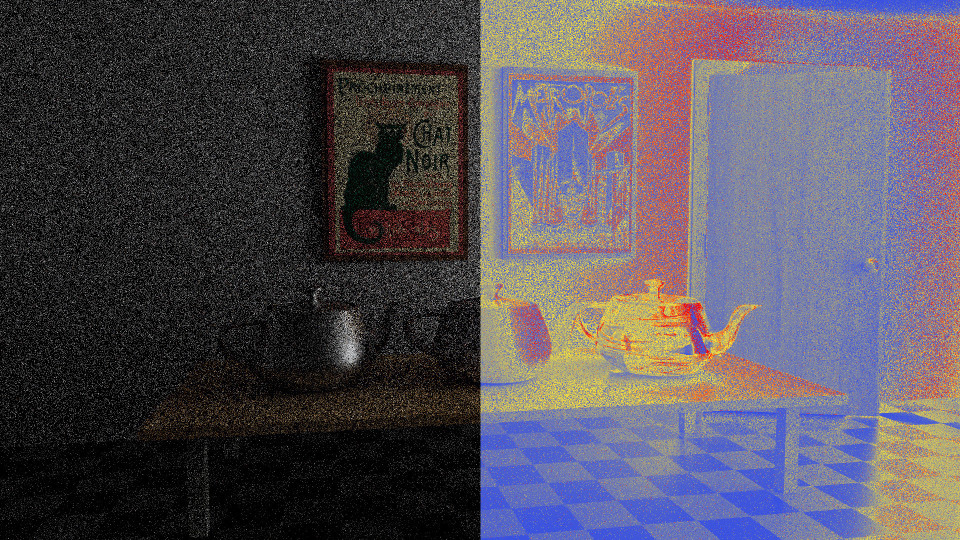}} & 	
		% 0.110337
		\multicolumn{1}{c}{\includegraphics[width=56.0mm]{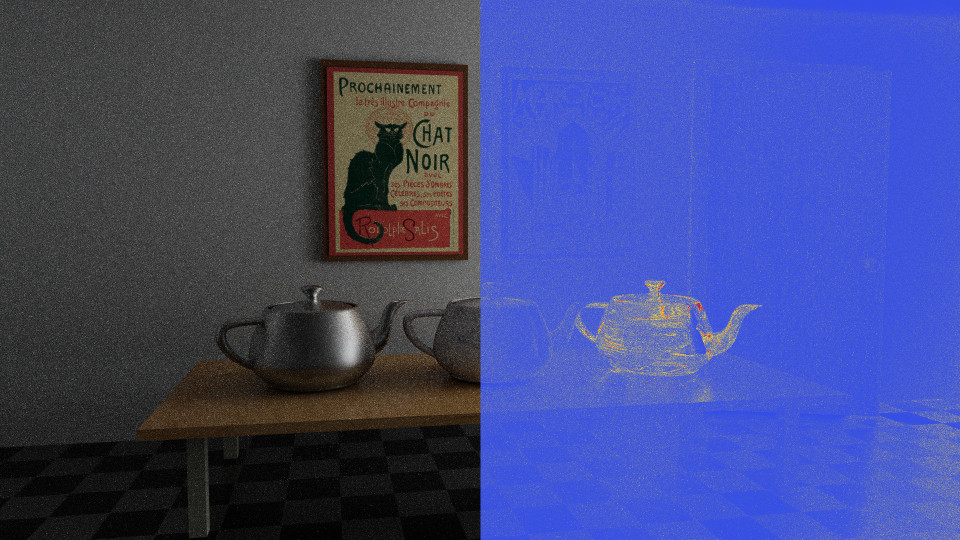}} 	
		\\ \hline
		&
		\tiny{RMSE 0.608228} &
		\tiny{RMSE 0.424356} &
		\tiny{RMSE 0.110337}
		\\ \hline
		\rotatebox{90}{PSTF-CV} &
		% 0.564476
		\multicolumn{1}{c}{\includegraphics[width=56.0mm]{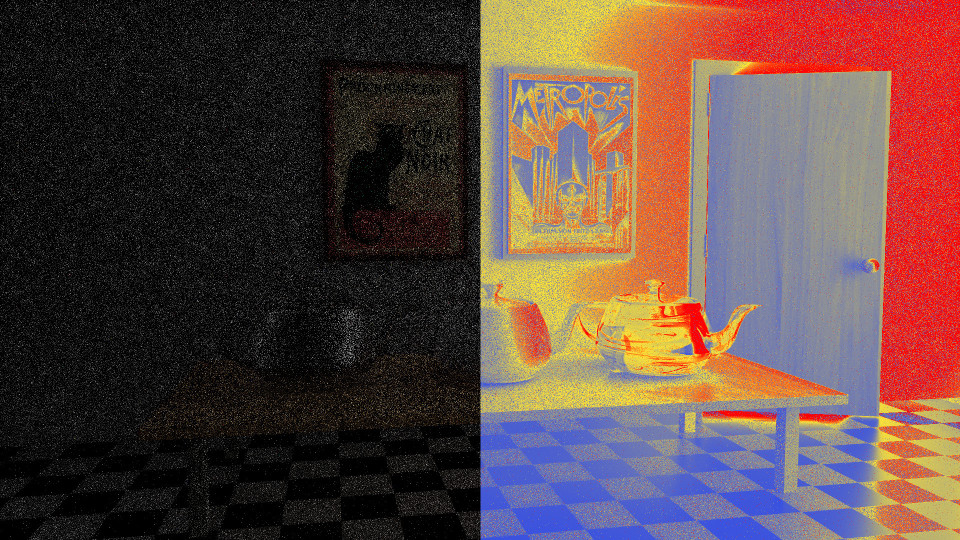}} &	
		% 0.367088
		\multicolumn{1}{c}{\includegraphics[width=56.0mm]{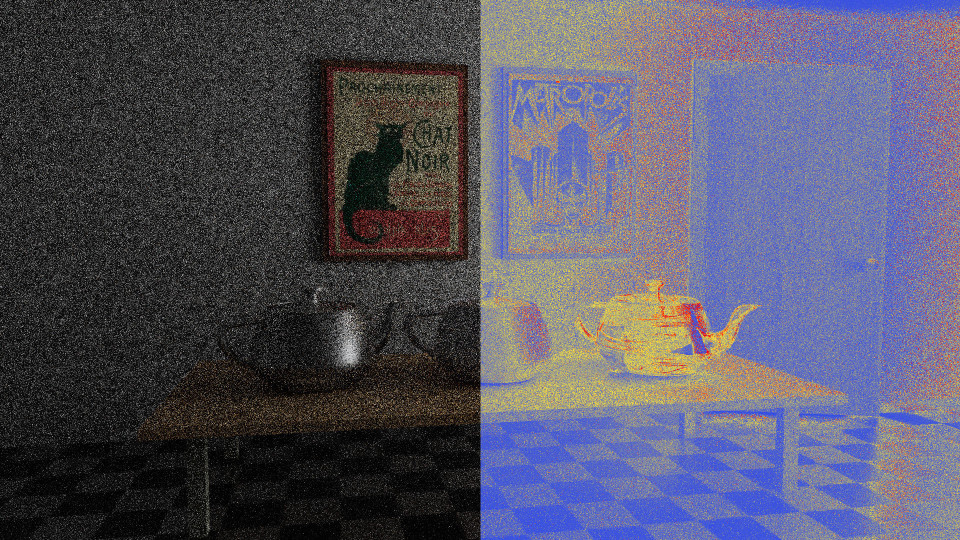}} &	
		% 0.126974
		\multicolumn{1}{c}{\includegraphics[width=56.0mm]{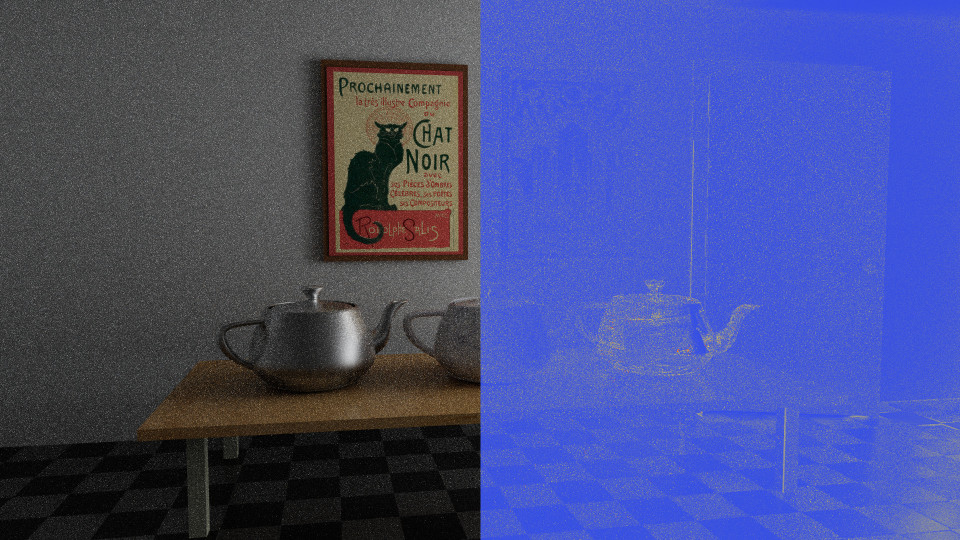}}	
		\\ \hline
		&
		\tiny{RMSE 0.564476} &
		\tiny{RMSE 0.367088} &
		\tiny{RMSE 0.126974}
		\\ \hline
		\rotatebox{90}{PSTF-IS-CV} &
		% 0.536593
		\multicolumn{1}{c}{\includegraphics[width=56.0mm]{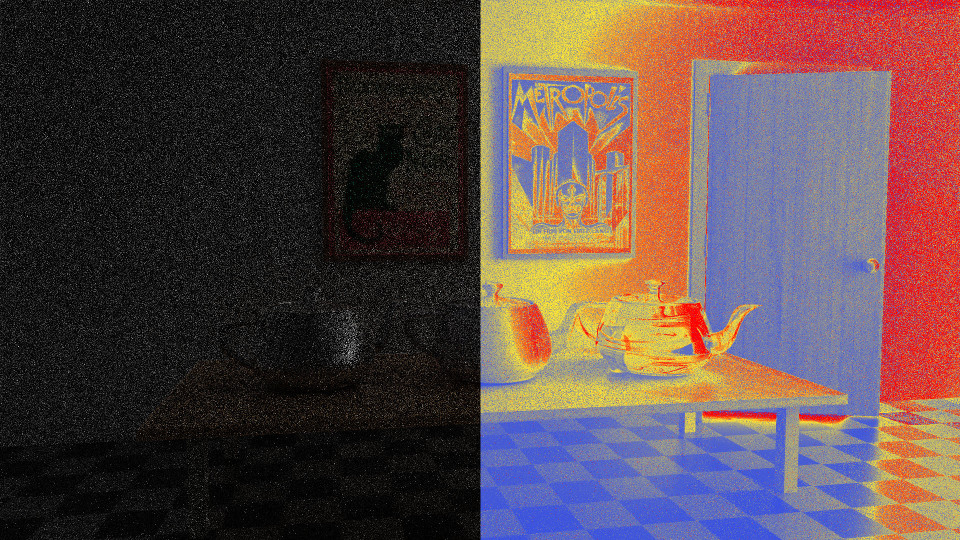}} &	
		% 0.374096
		\multicolumn{1}{c}{\includegraphics[width=56.0mm]{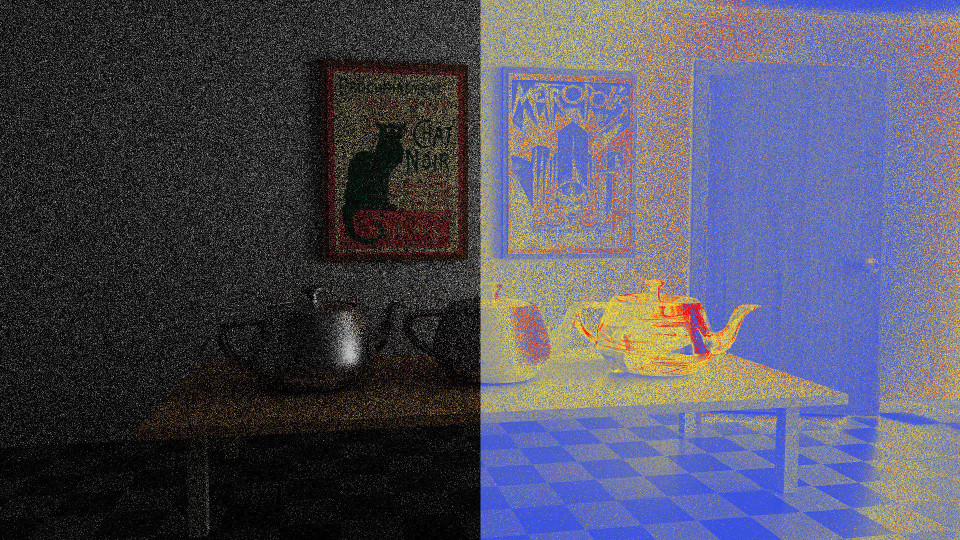}} &
		% 0.128888
		\multicolumn{1}{c}{\includegraphics[width=56.0mm]{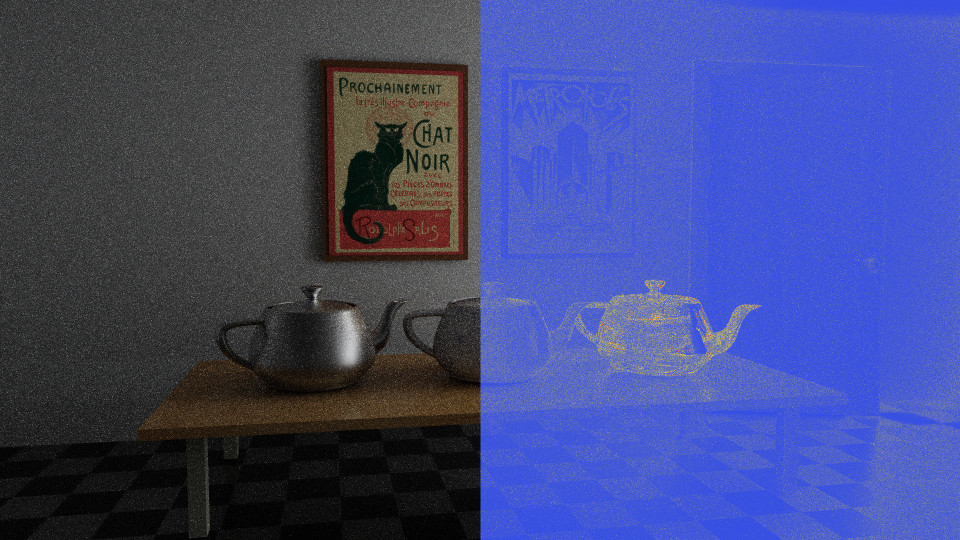}} 	
		\\ \hline
		&
		\tiny{RMSE 0.536593} &
		\tiny{RMSE 0.374096} &
		\tiny{RMSE 0.128888}
		\\ \hline
		\rotatebox{90}{PSTF-B} &
		% 0.508294
		\multicolumn{1}{c}{\includegraphics[width=56.0mm]{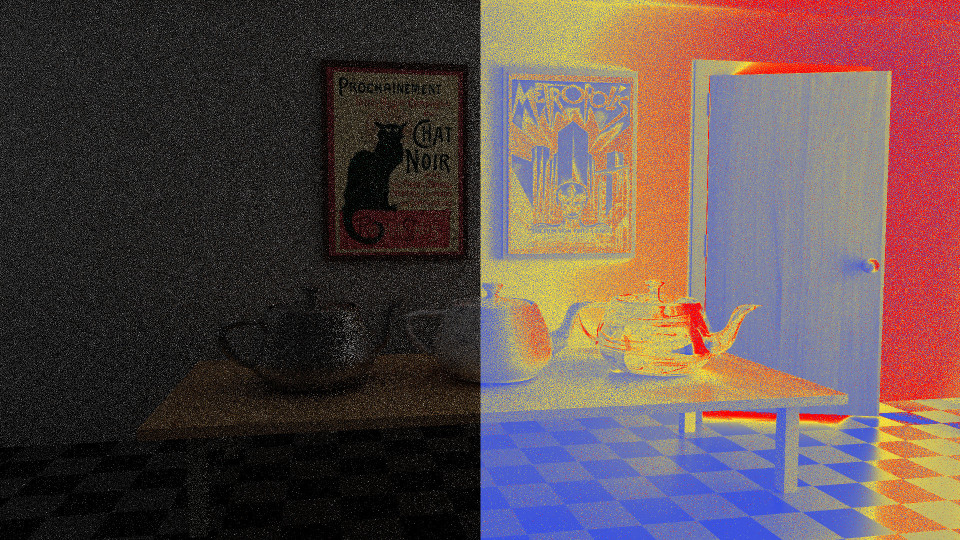}} &	
		% 0.215887
		\multicolumn{1}{c}{\includegraphics[width=56.0mm]{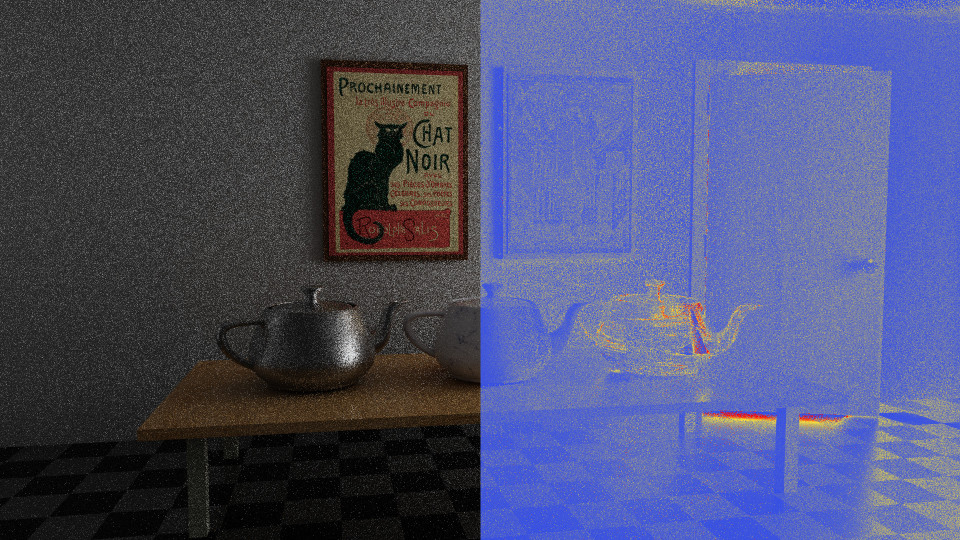}} &
		% 0.072084
		\multicolumn{1}{c}{\includegraphics[width=56.0mm]{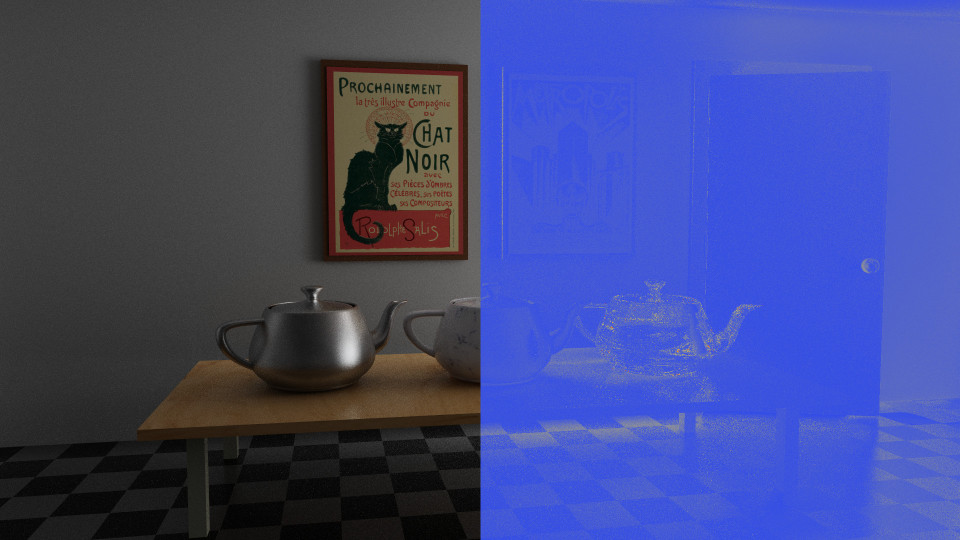}}
		\\ \hline
		&
		\tiny{RMSE 0.508294} &
		\tiny{RMSE 0.215887} &
		\tiny{RMSE 0.072084}
		\\ \hline
	\end{tabularx}
	\caption{
		\label{PSTF-comparisons}
		From top to bottom: 
		1. path tracing without NEE, 
		2. path tracing with NEE, 
		3. PSTF-IS, 
		4. PSTF-CV, 
		5. PSTF-IS-CV, 
		6. PSTF-B
		From left to right:
		1 spp for the first column,
		32 spp for the second column,
		and a same-time comparison for all the rows of the third column, after 60 seconds.
	}
	%\label{PSTF-comparisons}
	\addtolength{\tabcolsep}{4pt}
\end{table*}

\begin{figure*}
	\includegraphics[width=85.0mm]{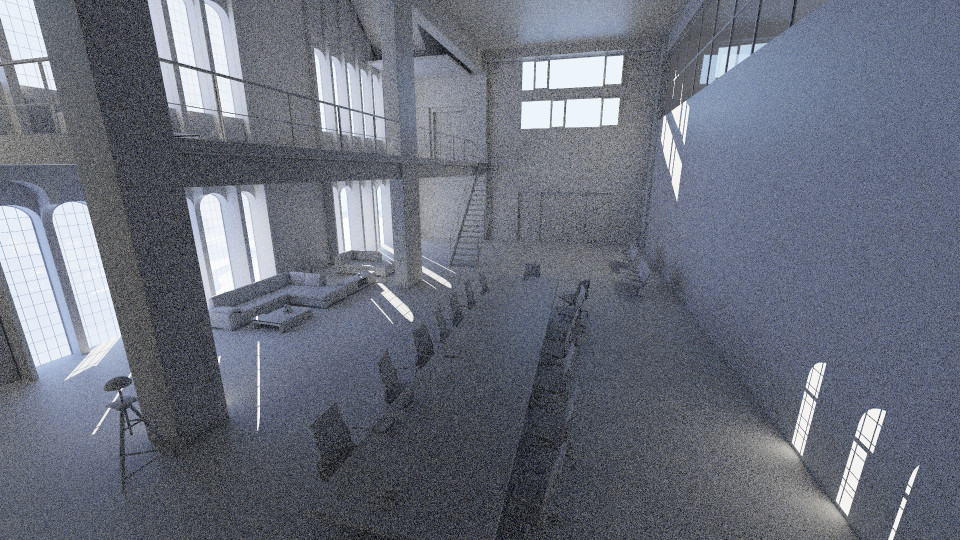}
	\includegraphics[width=85.0mm]{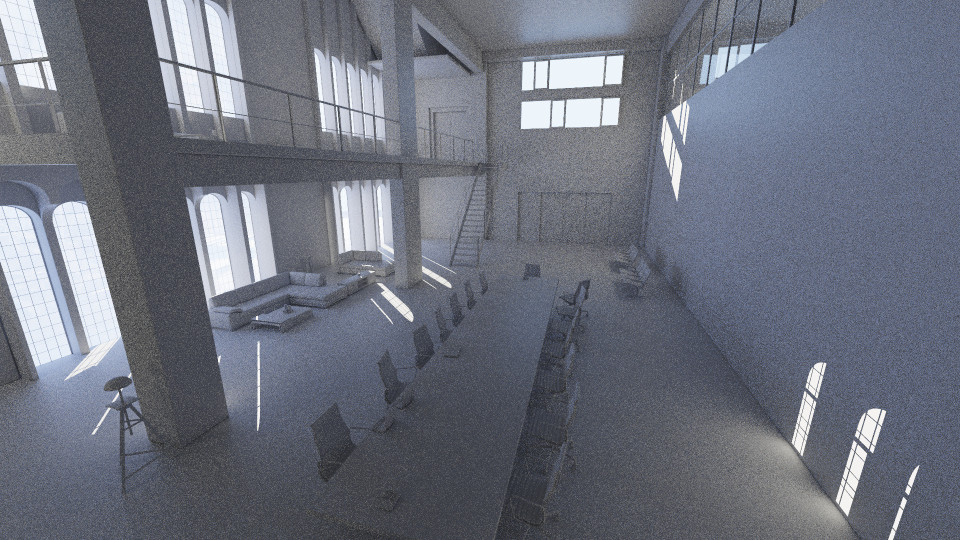}
	\caption{An equal-time comparison of PSF (left) and the biased PSTF estimator (right) applied at the first bounce, at 32spp, with the same pixel-sized spatial hashing. The progressive nature of PSTF allows for much quicker convergence.}
	\label{ArchInterior}
\end{figure*}

\RestyleAlgo{boxruled}
\begin{algorithm}%[frame=lines,label=inversion,caption={overall algorithm},captionpos=b]]
	\SetStartEndCondition{ }{}{}%
	\SetKwProg{Fn}{def}{\string:}{}
	\SetKwFunction{Range}{range}%%
	\SetKw{KwTo}{in}\SetKwFor{For}{for}{\string:}{}%
	\SetKwIF{If}{ElseIf}{Else}{if}{:}{elif}{else:}{}%
	\SetKwFor{While}{while}{:}{fintq}%
	\AlgoDontDisplayBlockMarkers\SetAlgoNoEnd\SetAlgoNoLine%
	
	\SetKwFunction{sampleCamera}{sample\_camera}
	\SetKwFunction{intersect}{intersect}
	\SetKwFunction{incrementCounters}{increment\_counters}
	\SetKwFunction{updateApprox}{update\_approx}
	\SetKwFunction{evalEmission}{eval\_emission}
	\SetKwFunction{sampleScattering}{scattering\_event}
	\SetKwFunction{nee}{next\_event\_estimation}
	{
		sample\_value = 0; \\
		sample\_weight = 1; \\
		
		\vspace{2mm}
		($x_0, \omega^i_0, p^\perp_0$) = \sampleCamera(); \\
		
		\vspace{2mm}
		// set the MIS weight to 1 (no other technique here) \\
		$w_0$ = 1; \\
		$f_0$ = 1; \\
		
		\vspace{2mm}
		\For {$j = 1$ \KwTo $\infty$}{
			$x_j$ = \intersect($x_{j-1}, \omega^i_{j-1}$); \\
			
			\vspace{2mm}
			$\omega^o_j$ = -$\omega^i_{j-1}$; \\
			
			\vspace{2mm}
			// increment the touched finite-elements' counters \\
			\incrementCounters($x_j, \omega^o_j$);
			
			\vspace{2mm}
			// update $\LoApproxNew$ at the previous vertex $(x_{j-1}, \omega^o_{j-1})$ \\
			// using information from the current \\
			\If {$j > 0$} {
				update\_value = \\
				\hspace{2mm} $\tilde{L}^{old}_o(x_j, \omega^o_j) \cdot f_{j-1} \cdot w_{j-1} / p^\perp_{j-1}$; \\
				
				\vspace{2mm}
				\updateApprox( \\
				\hspace{2mm} $x_{j-1}, \omega^o_{j-1}$, \\
				\hspace{2mm} update\_value ); \\
			}

			\vspace{2mm}
			// accumulate the local emission at this vertex \\
			sample\_value += sample\_weight $\cdot \, L_e(x_j, \omega^o_j)$; \\
			
			\vspace{2mm}
			// and accumulate it to $\LoApproxNew$ at this vertex \\
			update\_value = $L_e(x_j, \omega^o_j)$; \\
			\updateApprox( \\
				\hspace{2mm} $x_j, \omega^o_j$, \\
				\hspace{2mm} update\_value ); \\
			
			\vspace{2mm}
			// perform next event estimation \\
			nee\_value = \nee($x_j, \omega^o_j$); \\
			
			\vspace{2mm}
			// add its contribution to the output \\
			sample\_value += sample\_weight $\cdot$ nee\_value; \\
			
			\vspace{2mm}
			// sample a scattering event \\
			$(\omega^i_j, f_j, p^\perp_j, w_j)$ = \sampleScattering($x_{j}, \omega^o_j$); \\
			
			\vspace{2mm}
			// update the sample weight for this path \\
			sample\_weight = sample\_weight $\cdot f_j \cdot w_j / p^\perp_j$; 
		}
		
		\vspace{2mm}	
		\textbf{return} sample\_value;
	}
	\caption{\label{Alg1} basic PSTF path-tracing skeleton, tracking the field $\LoApprox$}
	%\label{Alg1}
\end{algorithm}

\RestyleAlgo{boxruled}
\begin{algorithm}%[frame=lines,label=inversion,caption={overall algorithm},captionpos=b]]
	\SetStartEndCondition{ }{}{}%
	\SetKwProg{Fn}{def}{\string:}{}
	\SetKwFunction{Range}{range}%%
	\SetKw{KwTo}{in}\SetKwFor{For}{for}{\string:}{}%
	\SetKwIF{If}{ElseIf}{Else}{if}{:}{elif}{else:}{}%
	\SetKwFor{While}{while}{:}{fintq}%
	\AlgoDontDisplayBlockMarkers\SetAlgoNoEnd\SetAlgoNoLine%
	
	\SetKwFunction{sampleCamera}{sample\_camera}
	\SetKwFunction{intersect}{intersect}
	\SetKwFunction{incrementCounters}{increment\_counters}
	\SetKwFunction{updateApprox}{update\_approx}
	\SetKwFunction{evalEmission}{eval\_emission}
	\SetKwFunction{sampleScattering}{scattering\_event}
	\SetKwFunction{nee}{next\_event\_estimation}
	{
		sample\_value = 0; \\
		sample\_weight = 1; \\
%		mis\_weight = 1; \\
		
		\vspace{2mm}
		($x_0, \omega^i_0, p^\perp_0$) = \sampleCamera(); \\
		
		\vspace{2mm}
		// set the MIS weight to 1 (no other technique here) \\
		$w_0$ = 1; \\
		$f_0$ = 1; \\
		
		\vspace{2mm}
		\For {$j = 1$ \KwTo $\infty$}{
			$x_j$ = \intersect($x_{j-1}, \omega^i_{j-1}$); \\
			
			\vspace{2mm}
			$\omega^o_j$ = -$\omega^i_{j-1}$; \\
			
			\vspace{2mm}
			// increment the touched finite-elements' counters \\
			\incrementCounters($x_j, \omega^o_j$);
			
			\vspace{2mm}
			// accumulate the local emission at this vertex \\
			sample\_value += sample\_weight $\cdot \, L_e(x_j, \omega^o_j)$; \\
			
			\vspace{2mm}
			// update $\LoeApproxNew$ at the previous vertex $(x_{j-1}, \omega^o_{j-1})$ \\
			// using information from the current \\
			\If {$j > 0$} {
				update\_value = \\
					\hspace{2mm} $[ L_e(x_j, \omega^o_j)$ \\
					\hspace{2mm} $\tilde{L}^{old}_{o \setminus e}(x_j, \omega^o_j) ] \cdot f_{j-1} \cdot w_{j-1} / p^\perp_{j-1}$; \\
				
				\vspace{2mm}
				\updateApprox( \\
				\hspace{2mm} $x_{j-1}, \omega^o_{j-1}$, \\
				\hspace{2mm} update\_value ); \\
			}
			
			\vspace{2mm}
			// perform next event estimation \\
			nee\_value = \nee($x_j, \omega^o_j$); \\
			
			\vspace{2mm}
			// add its contribution to the output \\
			sample\_value += sample\_weight $\cdot$ nee\_value; \\
			
			\vspace{2mm}
			// and (optionally) use it to update $\LoeApproxNew$ \\
			// at this vertex $(x_j, \omega^o_j)$ (recalling this \\
			// technique represents light transported \\
			// once, i.e. ${\bf T} L_e$) \\
			\updateApprox( \\
			\hspace{2mm} $x_j, \omega^o_j$, \\
			\hspace{2mm} nee\_value ); \\
			
			\vspace{2mm}
			// sample a scattering event \\
			$(\omega^i_j, f_j, p^\perp_j, w_j)$ = \sampleScattering($x_{j}, \omega^o_j$); \\
			
			\vspace{2mm}
			// update the sample weight for this path \\
			sample\_weight = sample\_weight $\cdot f_j \cdot w_j / p^\perp_j$; \\
%			mis\_weight = mis\_weight $\cdot w_j$; 
		}
		
		\vspace{2mm}	
		\textbf{return} sample\_value;
	}
	\caption{\label{Alg2} basic PSTF path-tracing skeleton, tracking the field $\LoeApprox$}
	%\label{Alg2}
\end{algorithm}

\section{Appendix}

Recall that a basis function $b_h \in \mathcal{H}$ is, like radiance, a function $b_h:\mathcal{R} \rightarrow \mathbb{R}$ on the ray space manifold $\mathcal{R} = \mathcal{M} \times S^2$, a product of the set of scene surfaces $\mathcal{M}$ and the sphere of directions $S^2$. The natural measure on ray space is the \emph{throughput measure}:
\begin{equation}
d\mu({\bf x}, \omega) = dA({\bf x}) \times d\sigma^{\perp}(\omega)
\end{equation}
In order to compute the throughput measure probability $P_T({\bf x}, \omega)$ of sampling a path landing on a point ${\bf x}$ from direction $-\omega$, we will start by considering the area probability $P_A({\bf x}, {\bf y})$ of sampling a path whose last two path vertices are, respectively, first ${\bf y}$ and finally ${\bf x}$. This is the product of two factors: the probability of sampling a path landing on ${\bf y}$, times the area probability $p({\bf x} | {\bf y})$ of sampling ${\bf x}$ given ${\bf y}$. The first factor can be obtained by integrating over all paths of all possible lengths $l$, so that we have:
\begin{eqnarray}
P_A({\bf x}, {\bf y}) &=& p({\bf x} | {\bf y}) \nonumber \\
&\cdot& \sum_{l=0}^\infty \int p({\xx}_0 \cdots {\xx}_l) p({\bf y}|{\xx}_{l}) dA({\xx}_{0}) \cdots dA({\xx}_{l}) \nonumber \\
\end{eqnarray}
Now, if we consider the fact that the point ${\bf y}$ can be deterministically obtained tracing a ray from ${\bf x}$ in direction $\omega$, i.e. ${\bf y} = h({\bf x}, \omega)$, we can convert between the area measure probability $P_A({\bf x}, {\bf y})$ and the throughput measure $P_T({\bf x}, \omega)$ with the formula:
\begin{equation}
P_T({\bf x}, \omega) = P_A({\bf x}, {\bf y}) \cdot \frac{ 1 }{ G({\bf x},{\bf y}) }
\end{equation}
where $G({\bf x},{\bf y})$ is the usual geometric throughput:
\begin{equation}
G({\bf x},{\bf y}) = \frac{ |\omega \cdot {\bf n_x}| |\omega \cdot {\bf n_y}|}{ |{\bf x} - {\bf y}|^2 }
\end{equation}
Finally, we can convert between the area probability $p({\bf x} | {\bf y})$ and the projected solid angle probability $p_{\bf y}^{\perp}(-\omega)$ of sampling the direction $-\omega$ at vertex ${\bf y}$ using the identity:
\begin{equation}
p({\bf x} | {\bf y}) = p_{\bf y}^{\perp}(-\omega) G({\bf x},{\bf y})
\label{AreaToSolidAngle}
\end{equation}
and combining the expressions we get:
\begin{eqnarray}
P_T({\bf x}, \omega) &=& p_{\bf y}^{\perp}(-\omega) \nonumber \\
&\cdot& \sum_{l=0}^\infty \int p({\xx}_0 \cdots {\xx}_l) p({\bf y}|{\xx}_{l}) dA({\xx}_{0}) \cdots dA({\xx}_{l}) \nonumber \\
\end{eqnarray}
with the position ${\bf y} = h({\bf x}, \omega)$.

%\begin{figure}
%	\includegraphics[width=84.0mm]{\picresdir/kd-vs-gmm.png}
%	\caption{An equal-time comparison graph of the RMSE of PSTF-IS using our spherical kd-trees vs spherical GMMs with 4 and 8 components respectively}
%	\label{KD-vs-GMM-fig}
%\end{figure}

%\begin{acks}
\paragraph{{\bf Aknowledgements}}
We thank Cem Cebenoyan and Aaron Lefohn for supporting this work and Jacob Munkberg and Pascal Gautron for carefully reviewing it and providing valuable feedbacks.
We would also like to acknowledge insightful discussions and related work by Alexander Keller, Nickolaus Binder, Sascha Fricke and Hannes Hergerth.
The \emph{staircase} scene is a modified version of a scene from the Rendering Resources archive \cite{resources16}.
The \emph{door ajar} scene is a modified version of the reconstruction performed by Miika Aittala, Samuli Laine, and Jaakko Lehtinen of the original scene by Eric Veach.
%\end{acks}

\fi

%\subsection{Parallel stepwise-EM}

%\input{../gmm}

\bibliographystyle{acmsiggraph}
\bibliography{main}

\begin{thebibliography}{\protect\citename{Vorba and K\v{r}iv\'{a}nek }2016}

\bibitem[\protect\citename{Bekaert et~al\mbox{.} }2003]{Bekaert:2003:CDE}
{\sc Bekaert, P., Slusallek, P., Cools, R., Havran, V., and Seidel, H.-P.}
\newblock 2003.
\newblock A custom designed density estimation method for light transport.
\newblock Research Report MPI-I-2003-4-004, Max-Planck-Institut f{\"u}r
  Informatik, Stuhlsatzenhausweg 85, 66123 Saarbr{\"u}cken, Germany, September.

\bibitem[\protect\citename{Binder et~al\mbox{.} }2018]{Binder:2018:FPS}
{\sc Binder, N., Fricke, S., and Keller, A.}
\newblock 2018.
\newblock Fast path space filtering by jittered spatial hashing.
\newblock In {\em ACM SIGGRAPH 2018 Talks}, ACM, New York, NY, USA, SIGGRAPH
  '18, 71:1--71:2.

\bibitem[\protect\citename{Bitterli et~al\mbox{.} }2017]{Bitterli:2017:RJM}
{\sc Bitterli, B., Jakob, W., Nov\'{a}k, J., and Jarosz, W.}
\newblock 2017.
\newblock Reversible jump metropolis light transport using inverse mappings.
\newblock {\em ACM Trans. Graph. 37}, 1 (Oct.), 1:1--1:12.

\bibitem[\protect\citename{Bitterli }2016]{resources16}
{\sc Bitterli, B.}, 2016.
\newblock Rendering resources.
\newblock https://benedikt-bitterli.me/resources/.

\bibitem[\protect\citename{{Chaos Group} }2008]{Vray:LC}
{\sc {Chaos Group}}, 2008.
\newblock {Light Cache GI}.
\newblock
  \url{https://docs.chaosgroup.com/display/VRAYSKETCHUP/Light+Cache+GI}.
\newblock [Online].

\bibitem[\protect\citename{Clarberg et~al\mbox{.} }2005]{Clarberg:2005:WIS}
{\sc Clarberg, P., Jarosz, W., Akenine-M\"{o}ller, T., and Jensen, H.~W.}
\newblock 2005.
\newblock Wavelet importance sampling: Efficiently evaluating products of
  complex functions.
\newblock In {\em ACM SIGGRAPH 2005 Papers}, ACM, New York, NY, USA, SIGGRAPH
  '05, 1166--1175.

\bibitem[\protect\citename{Dahm and Keller }2017]{Dahm:2017:LLT}
{\sc Dahm, K., and Keller, A.}
\newblock 2017.
\newblock Learning light transport the reinforced way.
\newblock In {\em ACM SIGGRAPH 2017 Talks}, ACM, New York, NY, USA, SIGGRAPH
  '17, 73:1--73:2.

\bibitem[\protect\citename{Hachisuka and Jensen }2009]{Hachisuka:2008:SPPM}
{\sc Hachisuka, T., and Jensen, H.~W.}
\newblock 2009.
\newblock Stochastic progressive photon mapping.
\newblock {\em {ACM} Trans. Graph. 28}, 5, 141.

\bibitem[\protect\citename{Hachisuka et~al\mbox{.} }2008]{Hachisuka:2008:PPM}
{\sc Hachisuka, T., Ogaki, S., and Jensen, H.~W.}
\newblock 2008.
\newblock Progressive photon mapping.
\newblock {\em {ACM} Trans. Graph. 27}, 5, 130.

\bibitem[\protect\citename{Hachisuka et~al\mbox{.} }2012]{Hachisuka:2012}
{\sc Hachisuka, T., Pantaleoni, J., and Jensen, H.~W.}
\newblock 2012.
\newblock A path space extension for robust light transport simulation.
\newblock {\em ACM Trans. Graph. 31}, 6 (Nov.), 191:1--191:10.

\bibitem[\protect\citename{Hanika et~al\mbox{.} }2015a]{Hanika:2015:MNE}
{\sc Hanika, J., Droske, M., and Fascione, L.}
\newblock 2015.
\newblock Manifold next event estimation.
\newblock {\em Comput. Graph. Forum 34}, 4 (July), 87--97.

\bibitem[\protect\citename{Hanika et~al\mbox{.} }2015b]{Hanika:2015:IHLST}
{\sc Hanika, J., Kaplanyan, A., and Dachsbacher, C.}
\newblock 2015.
\newblock Improved half vector space light transport.
\newblock {\em Computer Graphics Forum (Proceedings of Eurographics Symposium
  on Rendering) 34}, 4 (June), 65--74.

\bibitem[\protect\citename{He and Owen }2014]{He:2014:OMW}
{\sc He, H., and Owen, A.~B.}
\newblock 2014.
\newblock Optimal mixture weights in multiple importance.
\newblock Research report, November.

\bibitem[\protect\citename{Herholz et~al\mbox{.} }2016]{Herholz:2016:PIS}
{\sc Herholz, S., Elek, O., Vorba, J., Lensch, H., and K\v{r}iv\'{a}nek, J.}
\newblock 2016.
\newblock Product importance sampling for light transport path guiding.
\newblock {\em Comput. Graph. Forum 35}, 4 (July), 67--77.

\bibitem[\protect\citename{Ivo et~al\mbox{.} }2019]{Kondapaneni:2019:OMIS}
{\sc Ivo, K., V{\'{e}}voda, P., Grittmann, P., Sk{\v{r}}ivan, T., Slusallek,
  P., and K{\v{r}}iv{\'{a}}nek, J.}
\newblock 2019.
\newblock Optimal multiple importance sampling.
\newblock {\em ACM Transactions on Graphics (Proceedings of SIGGRAPH 2019) 38},
  4 (July), 37:1--37:14.

\bibitem[\protect\citename{Jakob and Marschner }2012]{Jakob:2012:ME}
{\sc Jakob, W., and Marschner, S.}
\newblock 2012.
\newblock Manifold exploration: A markov chain monte carlo technique for
  rendering scenes with difficult specular transport.
\newblock {\em ACM Trans. Graph. 31}, 4 (July), 58:1--58:13.

\bibitem[\protect\citename{Kajiya }1986]{Kajiya:1986:RE}
{\sc Kajiya, J.~T.}
\newblock 1986.
\newblock The rendering equation.
\newblock In {\em Proceedings of the 13th Annual Conference on Computer
  Graphics and Interactive Techniques}, ACM, New York, NY, USA, SIGGRAPH '86,
  143--150.

\bibitem[\protect\citename{Kaplanyan et~al\mbox{.} }2014]{Kaplanyan:2014:HSLT}
{\sc Kaplanyan, A.~S., Hanika, J., and Dachsbacher, C.}
\newblock 2014.
\newblock The natural-constraint representation of the path space for efficient
  light transport simulation.
\newblock {\em ACM Transactions on Graphics (Proc. SIGGRAPH) 33}, 4.

\bibitem[\protect\citename{Karl{\'i}k et~al\mbox{.} }2019]{Karlik:2019:MISC}
{\sc Karl{\'i}k, O., {\v{S}}ik, M., V{\'e}voda, P., Sk{\v{r}}ivan, T., and
  K{\v{r}}iv{\'{a}}nek, J.}
\newblock 2019.
\newblock Mis compensation: Optimizing sampling techniques in multiple
  importance sampling.
\newblock {\em ACM Trans. Graph. (SIGGRAPH Asia 2019) 38}, 6.

\bibitem[\protect\citename{Kelemen et~al\mbox{.} }2002]{Kelemen:2002}
{\sc Kelemen, C., Szirmay-Kalos, L., Antal, G., and Csonka, F.}
\newblock 2002.
\newblock A simple and robust mutation strategy for the {M}etropolis light
  transport algorithm.
\newblock In {\em Computer Graphics Forum}, 531--540.

\bibitem[\protect\citename{Keller et~al\mbox{.} }2014]{Keller:2014:PSF}
{\sc Keller, A., Dahm, K., and Binder, N.}
\newblock 2014.
\newblock Path space filtering.
\newblock In {\em ACM SIGGRAPH 2014 Talks}, ACM, New York, NY, USA, SIGGRAPH
  '14, 68:1--68:1.

\bibitem[\protect\citename{Lafortune and Willems }2016]{Lafortune:1995:5D}
{\sc Lafortune, E., and Willems, Y.}
\newblock 2016.
\newblock A 5d tree to reduce the variance of monte carlo ray tracing.
\newblock {\em Rendering Techniques 1995 (Proceedings of the Sixth Eurographics
  Workshop on Rendering)\/}, 11--20.

\bibitem[\protect\citename{Li et~al\mbox{.} }2013]{Li:2013:TIS}
{\sc Li, W., Tan, Z., and Chen, R.}
\newblock 2013.
\newblock Two-stage importance sampling with mixture proposals.
\newblock {\em Journal of the American Statistical Association 108}, 504,
  1350--1365.

\bibitem[\protect\citename{M\"{u}ller et~al\mbox{.} }2017]{Muller:2017:PPG}
{\sc M\"{u}ller, T., Gross, M., and Nov\'{a}k, J.}
\newblock 2017.
\newblock Practical path guiding for efficient light-transport simulation.
\newblock {\em Comput. Graph. Forum 36}, 4 (July), 91--100.

\bibitem[\protect\citename{M\"{u}ller et~al\mbox{.} }2019]{Muller:2019:NIS}
{\sc M\"{u}ller, T., Mcwilliams, B., Rousselle, F., Gross, M., and Nov\'{a}k,
  J.}
\newblock 2019.
\newblock Neural importance sampling.
\newblock {\em ACM Trans. Graph. 38}, 5 (Oct.), 145:1--145:19.

\bibitem[\protect\citename{Owen and Zhou }2000]{Owen:2000:SAE}
{\sc Owen, A.~B., and Zhou, Y.}
\newblock 2000.
\newblock Safe and effective importance sampling.

\bibitem[\protect\citename{Pantaleoni }2017]{Pantaleoni:2017:CML}
{\sc Pantaleoni, J.}
\newblock 2017.
\newblock Charted metropolis light transport.
\newblock {\em ACM Trans. Graph. 36}, 4 (July), 75:1--75:14.

\bibitem[\protect\citename{Pantaleoni }2019]{Pantaleoni:2019:RLL}
{\sc Pantaleoni, J.}
\newblock 2019.
\newblock {Importance Sampling of Many Lights with Reinforcement Lightcuts
  Learning}.
\newblock {\em {arXiv:1911.10217}\/} (Dec.).

\bibitem[\protect\citename{Sbert and Elvira }2019]{Sbert:2019:GMIS}
{\sc Sbert, M., and Elvira, V.}
\newblock 2019.
\newblock {Generalizing the Balance Heuristic Estimator in Multiple Importance
  Sampling}.
\newblock {\em {arXiv:1903.11908}\/} (Mar.).

\bibitem[\protect\citename{Veach and Guibas }1997]{Veach:1997:MLT}
{\sc Veach, E., and Guibas, L.~J.}
\newblock 1997.
\newblock Metropolis light transport.
\newblock In {\em Proceedings of the 24th Annual Conference on Computer
  Graphics and Interactive Techniques}, ACM Press/Addison-Wesley Publishing
  Co., New York, NY, USA, SIGGRAPH '97, 65--76.

\bibitem[\protect\citename{Veach }1997]{Veach:PHD}
{\sc Veach, E.}
\newblock 1997.
\newblock {\em Robust Monte Carlo Methods for Light Transport Simulation}.
\newblock PhD thesis, Stanford University.

\bibitem[\protect\citename{Vorba and K\v{r}iv\'{a}nek }2016]{Vorba:2016:ARR}
{\sc Vorba, J., and K\v{r}iv\'{a}nek, J.}
\newblock 2016.
\newblock Adjoint-driven russian roulette and splitting in light transport
  simulation.
\newblock {\em ACM Trans. Graph. 35}, 4 (July), 42:1--42:11.

\bibitem[\protect\citename{Vorba et~al\mbox{.} }2014]{Vorba:2014:OLP}
{\sc Vorba, J., Karl{\'i}k, O., {\v{S}}ik, M., Ritschel, T., and
  K{\v{r}}iv{\'{a}}nek, J.}
\newblock 2014.
\newblock On-line learning of parametric mixture models for light transport
  simulation.
\newblock {\em ACM Transactions on Graphics (Proceedings of SIGGRAPH 2014) 33},
  4 (aug).

\bibitem[\protect\citename{Vorba et~al\mbox{.} }2019]{vorba19guiding}
{\sc Vorba, J., Hanika, J., Herholz, S., M\"{u}ller, T., K\v{r}iv\'{a}nek, J.,
  and Keller, A.}
\newblock 2019.
\newblock Path tracing in production.
\newblock In {\em ACM SIGGRAPH Courses}, ACM, New York, NY, USA, 18:1--18:77.

\end{thebibliography}

\end{document}